# Hall probes: physics and application to magnetometry

*S. Sanfilippo*
Paul Scherrer Institut, Villigen, Switzerland

*"What I do teaches me what I am looking for"*
Pierre Soulages, 1953, retrospective of the 'painter of black and light', 14 October 2009 – 8 March 2010 (Pompidou Centre, Paris)

**Abstract**
This lecture aims to present an overview of the properties of Hall effect devices. Descriptions of the Hall phenomenon, a review of the Hall effect device characteristics and of the various types of probes are presented. Particular attention is paid to the recent development of three-axis sensors and the related techniques to cancel the offsets and the planar Hall effect. The lecture introduces the delicate problem of the calibration of a three-dimensional sensor and ends with a section devoted to magnetic measurements in conventional beam line magnets and undulators.

## 1    Introduction

The measurement of magnetic fields plays a key role in various areas. The mastering of the magnetic field intensity allows the construction of alternators and of electrical motors, the storage of information on magnetic tapes or hard disks, and the study of phenomena in condensed matter physics like the flux distribution in superconductors. Magnetic fields of very strong intensity are used in particle accelerators or in Tokomaks to guide and focus beams with high energy and make them collide. Field measurements are also omnipresent in the domain of medicine to detect heart and brain activity. The Hall effect is an ideal magnetic field sensing technology. Discovered in 1879 by a professor of the Johns Hopkins University in the USA, E. H. Hall [1], its application remained confined to laboratory experiments until the 1950s as a tool for measuring the magnetic field or exploring the basic electronic properties of a solid. With the mass production of semiconductors and the development of the associated electronics, it became progressively feasible to combine a Hall effect sensing element with electronics in a single integrated circuit. This changed the end objective of the Hall effect devices. From the simple measurement of the magnetic field, they became a key element in the sensing of pressure, current, position or temperature via magnetic field detection. Hall sensors can be found nowadays in many products ranging from computers to automobiles, aircraft and medical equipment. An overview of the Hall effect devices used as magnetic sensors can be found, for example, in the bibliographic items [B1]–[B6].

The object of this lecture is to describe the principle of the Hall effect, the basic characteristics of a Hall device, the advantages and the limitations of this technique in the field of magnetic measurements and magnetic field mapping. Obviously, given the vastness of the domain of application, this cannot be an exhaustive review of Hall probe measurements. The progress on the state of the art and on the domain of application is a typical topic of discussions in conferences [2], workshops [3] and international journals [4]. We will focus on the case of a three dimensional axis Hall device with an overview of the recent progress that has been accomplished to minimize the planar Hall effect and the offset. The delicate problem of the calibration of a three dimensional Hall device is briefly introduced. The course ends with two examples of magnetic measurements with a Hall probe: in beam line magnets and in undulators.

## 2 The Hall generator

### 2.1 The Hall generator

A Hall generator consists of a sub-millimetre (semi)conductor element, the active part, typically ranging from 0.01 mm to 0.1 mm which is sensitive to magnetic field. It is equipped with two current electrodes (biasing contacts) to conduct the current *I* into the conducting plate and with two voltage electrodes to measure the Hall voltage $V_H$ (sensing contact). The Hall element is affixed to a ceramic substrate which provides improved mechanical support and thermal stability and encapsulated. Figure 1 shows a typical Hall generator construction. In its basic form, a current supplied by a highly stable AC or DC current source feeds two terminals and the Hall voltage is read through the other two terminals by a high impedance voltmeter or amplified and conditioned by acquisition electronics. This device, combined with associated electronics, is used to measure the magnetic field based on the Hall effect and is commonly called Hall sensor. It is part of the category of transducers that convert a non-electrical energy (the magnetic field) into an electrical one (the Hall voltage). In this lecture the following definitions will be used [5]: the denomination "Hall generator" or "Hall device" will refer to a four terminal device containing a Hall element and exploiting the Hall effect. The Hall generator mounted on a holder will be called "Hall probe". The term "Hall plate" reflects a conventional form of the Hall device. The denomination "Hall sensor" will be used to stress the application of a Hall generator as magnetic sensor. The combination of a Hall device with some electronic circuitry is called "Integrated Hall magnetic sensor" or "IC Hall sensor".

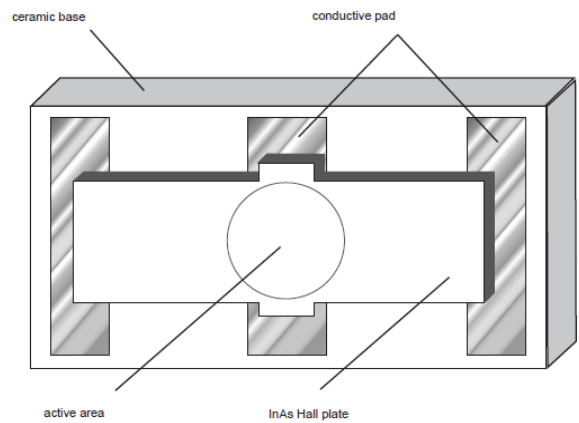

**Fig. 1:** Typical Hall generator construction (picture from Ref. [6])

### 2.2 The Hall effect (simplified approach)

The Hall effect is a galvanomagnetic phenomenon caused by the current flow deviation in presence of a magnetic field component perpendicular to the flow direction due to the Lorentz force. The response of this phenomenon, i.e., the voltage produced, was found to be proportional to the magnetic field intensity *B*. While the effect was originally detected in a thin metallic (gold) foil, modern Hall elements are nowadays mostly produced in semiconductors such as silicon (Si), indium arsenide (InAs), gallium arsenide (GaAs) or indium antimonide (InSb). This section describes the Hall effect in its simplest and classic form, where a long current-carrying strip is exposed to a magnetic field as shown in Fig. 2. The Hall generator consists of a thin semiconductor plate of dimensions (*l*, *w*, *t*), equipped with four contacts. The control current *I* flows through the biasing contacts CC while the sense contacts, S1S2, are used for the measurement of the Hall voltage. The magnetic field **B** is normal to the plate or making an angle γ (generally small) with respect to the normal of the strip. We assume that:

---

NB: The physical quantities in bold refer to vectors.

- the material is strongly extrinsic n-type, i.e., the majority carriers are electrons (particle charge e);
- the role of the holes is neglected;
- the velocity is the same for all the electrons (smooth drift approximation);
- the effect of the thermal agitation is neglected;
- the material is plate like and isotropic;
- the strip is infinitely long;
- the Hall contacts are extremely small.

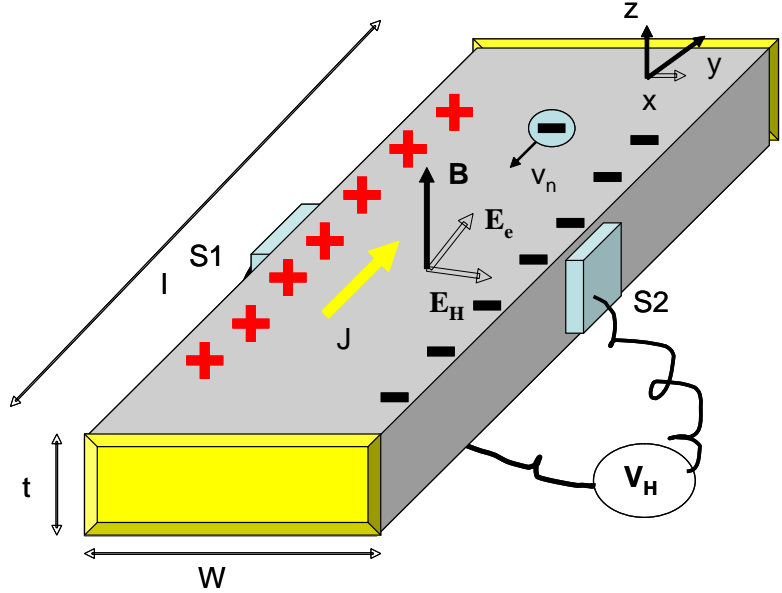

**Fig. 2:** Sketch of a Hall generator. An applied magnetic field **B** (normal to the plate in that case) is applied and a current *I* (current density *J*) flows in the *y* direction through the CC contacts. The output voltage is measured through the S1S2 contacts

When no magnetic field is applied, the electrons move in the longitudinal direction (*y* direction) parallel to the longitudinal external electrical field $\mathbf{E_e}$. As soon as a magnetic field is applied, the trajectory is deflected and the electrons with a velocity $\mathbf{v_n}$ move under the action of the Lorentz force $\mathbf{F_L}$ in response to the magnetic field **B** and the external electric field $\mathbf{E_e}$. $\mathbf{F_L}$ is given by

$$\mathbf{F_L} = e\,(\mathbf{E_e} + \mathbf{v_n} \times \mathbf{B}) \quad e = -q \text{ for electrons}. \tag{1}$$

The electrons are pressed by the magnetic force towards one edge of the strip. Consequently, the concentration of the electrons at the other edge is decreasing. To re-establish the charge repartition, an electric field $\mathbf{E_H}$ is created between the two edges and moves the carriers so as to decrease the excess of charges. Equilibrium is reached when the magnetic force pushing the carriers aside corresponds to this reacting electric force trying to push them back to the middle. The equilibrium is expressed by

$$\mathbf{E_H} + \mathbf{v_n} \times \mathbf{B} = 0. \tag{2}$$

At this point, the electrons will again move in the longitudinal direction. This leads to the definition of the Hall field $\mathbf{E_H}$ that is only a function of the velocity of the electrons and of the applied magnetic field. $\mathbf{E_H}$ is given by

$$\mathbf{E_H} = -\mathbf{v_n} \times \mathbf{B}. \tag{3}$$

A measurable transverse voltage $V_H$ called the Hall voltage is produced between the two edges of the strip. This represents the most tangible effect associated with the Hall effect. We will express the Hall voltage as a function of the current and the magnetic field.

The velocity of the electron is related to its mobility $\mu_n$ and the external electric field by

$$\mathbf{v}_n = -\mu_n \mathbf{E}_e . \tag{4}$$

The current density $J$ is related to the electric field by Ohm's law:

$$\mathbf{J} = \sigma \mathbf{E}_e . \tag{5}$$

The material conductivity $\sigma$ is expressed by

$$\sigma = n\, q\, \mu_n , \quad \text{where } n \text{ is the density of electrons} . \tag{6}$$

The associated current density for electrons is therefore given by

$$\mathbf{J} = n\, q\, \mu_n \mathbf{E}_e = -n\, q\, \mathbf{v}_n . \tag{7}$$

The Hall voltage $V_H$ appearing across the two sense electrodes (S1S2) can be calculated along the width of the strip by

$$V_H = \int_{S1S2} \vec{E}_H\, \vec{dl} = -\frac{J}{nq} B\, w . \tag{8}$$

$J$ is related to the biasing current $I$ by $\quad J = I / w\, t \quad$ (see geometry in Fig. 2) . $\tag{9}$

For a homogeneous, isotropic, rectangular and infinitely long Hall generator, the Hall voltage can be related to the current and the magnetic field by combining Eqs. (8) and (9):

$$V_H = \frac{R_H}{t} I\, B . \tag{10}$$

$R_H$ is named the Hall coefficient and is given **for the electrons** by

$$R_H = -1/n\, q = -\mu_n / \sigma . \tag{11}$$

More generally, if **B** is slightly tilted with respect to the normal, making an angle γ with respect to the normal of the strip, the Hall voltage $V_H$ is given by

$$V_H = \frac{R_H}{t} I\, B \cos\gamma . \tag{12}$$

$R_H$ is the parameter that gauges the magnitude of the Hall effect.

Remark 1: The Hall voltage $V_H$ depends linearly on

– the charge carrier velocity;
– the magnetic field intensity $B$;
– the distance between the sense contacts.

Remark 2: The Hall constant $R_H$ is related to the density of the electrons inside the support material and the mobility of the material [Eq. (11)]. This leads to one direct application of the Hall effect. The measurement of the Hall coefficient of materials with a known conductivity will indicate directly the mobility of the carriers responsible for the conductance.

- The minus sign above is obtained for electrons, i.e., negative charges.
- If positively charged carriers would be involved, the sign of the Hall constant would be positive.

Remark 3: In addition to the carrier mobility and the geometry (see Section 2.2), the sample thickness is of crucial importance. It determines the amplitude of the output signal following the relationship

$$V_H \propto 1/t. \qquad (13)$$

Remark 4: Hall angle $\Theta_H$

The total electrical field **E** is the sum of the external field **E**$_e$ and of the Hall field **E**$_H$ which are not parallel. The Hall angle $\Theta_H$ is defined as the angle between **E**$_e$ and the total electric field **E** (see Fig. 3). The angle value depends on magnetic field intensity and on the mobility of the charge carrier according to Eq. (14). The sign of the Hall angle coincides with the sign of the charge carrier.

$$tg\,\Theta_H = \frac{|E_H|}{|E_e|} = -\mu_n\,B \qquad (14)$$

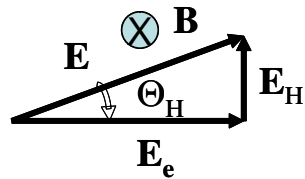

**Fig. 3:** Definition of the Hall angle in the case of a n-type semiconductor

## 2.3 Hall element geometry and choice of the material

### 2.3.1 Geometry

For real Hall generators, the actual dimensions and the finite size of the Hall contacts have to be taken into account. These two dependences can be regrouped in a dimensionless geometrical correction factor $G$ ranging from 0 to 1 that multiplies the Hall voltage formula in Eq. (10).

$$V_H = G\,(l/w,\,\Theta_H,\,B)\,R_H\,I\,B\,/\,t\,. \qquad (15)$$

It can be demonstrated that this factor depends on the geometry, on the dimensions of the active part, on the Hall angle $\Theta_H$, and on the magnetic field. This factor determines the reduction of the Hall voltage due to the imperfect bias current confinement. It depends mainly on the ratio $l/w$. Figure 4 shows some typical shapes of Hall generators. The bridge shape (a) is a good approximation of a long Hall device and allows relatively large contacts. The parasitic effects from the contact resistance and the heating can therefore be minimized. The cruciform (b) and the Van der Paw (d) shapes are of particular interest. They offer the advantages that a high geometrical correction factor $G \sim 1$ can be achieved, the geometry is simple and compact and the shape is invariant for a rotation through $\pi/2$. The current and the sense contacts could be switched without changing the global symmetry. This four-fold symmetry is favourable for the application of offset voltage compensation techniques like the spinning current technique (see Section 4.4). This geometry also allows a better definition of the active part centre.

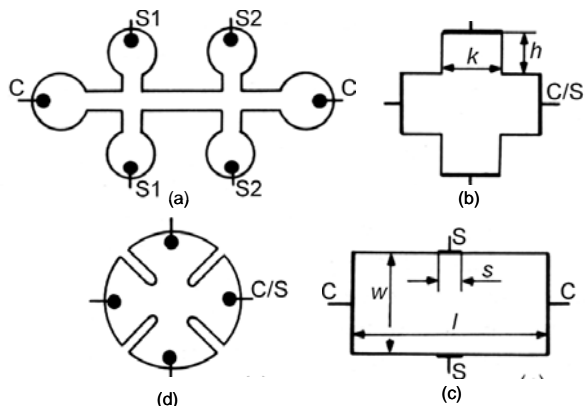

**Fig. 4:** Various shapes of Hall elements. C are the current contacts, S the sense contacts and C/S indicates that the two types of contacts are interchangeable [from Ref. B1, p.194]

### 2.3.2 Material

The choice of the proper material for the Hall element is of crucial importance. From Eq. (11), one can conclude that the Hall effect will be favoured in materials with high mobility and also low conductivity. Metals show low mobility and high conductivity and are therefore not a good choice. The sensors are usually prepared from n-type semiconductors where the dominant charge carriers are the electrons having much higher mobility than holes. Suitable candidates are single elements like Si and III-V compounds like InSb, InAs and GaAs. The III-V compounds combine high carrier mobility and reasonable value of conductivity. Silicon has moderate electron mobility but is compatible with the integrated circuit technology. This makes this material very attractive for the realization of Hall sensor chips. Table 1 gives the values of the semiconducting gap $E_g$ and the mobility at room temperature of the various semiconductors used for Hall generators. $R_H$ is also calculated for a given doping density.

**Table 1**: Gap and mobility of semiconductors at 300 K used to build a Hall plate (from Ref. [7]). $R_H$ is calculated for a given level of doping

| Material | $E_g$ [eV] | $\mu_n$ [cm$^{-2}$V$^{-1}$s$^{-1}$] | $n$ [cm$^{-3}$] | $R_H$ [cm$^3$C$^{-1}$] |
|---|---|---|---|---|
| Si | 1.12 | 15 00 | 2.5 10$^{15}$ | 2.5 × 10$^3$ |
| InSb | 0.17 | 80 000 | 9 10$^{16}$ | 70 |
| InAs | 0.36 | 33 000 | 5 10$^{16}$ | 125 |
| GaAs | 1.42 | 85 00 | 1.45 10$^{15}$ | 2.1 × 10$^3$ |

## 2.4 Basic characteristics and parasitic effects on Hall probes

The quality of a Hall probe can be appreciated with the following criteria, see Refs. [B7, B8, B9]:

- the sensitivity to the magnetic field;
- the low zero-field offset and its small drift in time;
- low temperature-dependence coefficient;
- the small linearity error of the Hall voltage (% of non-linearity);
- the minimization of the thermal, electrical and frequency noise;

- very low power consumption;
- possibility to operate in a large temperature range (low temperature applications);
- and also … the cost.

In the following, some of those basic features of a Hall sensor will be developed.

### 2.4.1 *Magnetosensitivity*

The magnetosensitivity (we will use the term of sensitivity) expresses the response of the output voltage to a magnetic field. The (absolute) sensitivity $S$ can be formulated by

$$S = \frac{V_H}{B_\perp} \quad [\text{V/T}] \qquad B_\perp \text{ represents the normal component of } \mathbf{B} \text{ to the Hall plate.} \qquad (16)$$

Two other figures of merit can also be defined:

– the supply current-related sensitivity $S_I$ in Volts per unit field, per unit of bias current $I_{HP}$ ;

– the supply voltage-related sensitivity $S_V$ in volts per unit field, per unit of bias voltage mostly employed when the additional input is fed by a supply voltage $V_s$

$$S_I = \frac{S}{I_{HP}} \quad [\text{V/AT}] \quad \text{and} \quad S_V = \frac{S}{V_s} \quad [\text{V/VT}] . \qquad (17)$$

Modern devices $S$ show values typically ranging between 50 mV/T and 1 V/T at the nominal biasing current. The sensitivity drifts with time which necessitates periodic calibration, $V_H = f(B)$, of the Hall probe (see Section 5).

### 2.4.2 *Offset voltage*

The offset voltage $V_{off}$ is a parasitic output voltage that appears in the absence of a magnetic field. Some origins of this voltage are

– a structural asymmetry of the active part (errors in geometry, non uniform doping density, contact resistance, etc.);

– alignment errors of the sense contacts, one being further upstream and the other one downstream with respect to the bias current [see illustration in Fig. 5 (left)].

The offset changes with time. The drift of the offset voltage originates from

– the ageing of the semiconductor that plays an important role for cryogenic Hall probes. Thermal cycles between 4.4 K and 300 K will cause damage in the active area (microcracks), changes in contacts, resin glue and package. The mechanical stress that appears will modify the repartition of the electron population in a many-valley semiconductor like n-type silicon. This evolution of concentration will change the Hall coefficient and hence the Hall voltage.

– The diffusion of the impurities from the contact into the active area, modifying the mobility and thus the Hall coefficient. This effect can be limited using a cruciform geometry.

It is practical to use the analogy of the resistive bridge to model the presence of the offset. Ideally the four resistors are equal. A resistance variation in one or more legs leads to bridge asymmetry resulting in the appearance of an offset [Fig. 5 (right)].

In conventional 'plate-like' Hall probes, the offset voltage is in most cases small, in the range of a microvolt. In some integrated circuit Hall sensors, however, it can reach a value of hundreds of microvolts. It is linearly dependent on the biasing current and is temperature dependent. The measured output voltage is therefore the sum of the offset and of the Hall voltages and the two contributions cannot be distinguished when $B \neq 0$. The offset can be measured in a zero-field gauss chamber or compensated by means of external circuits. Compensation techniques are presented in Section 4.4.1.

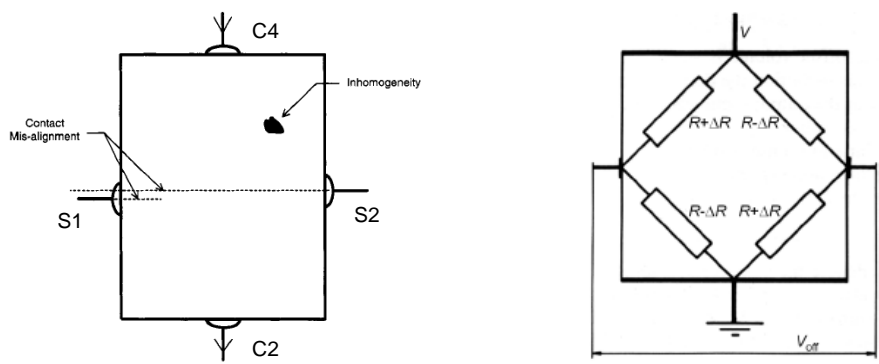

**Fig. 5:** Left: Offset resulting from the misalignment of sense contacts and from inhomogeneities in the material [B2, p.13]. Right: Analogy with an unbalanced bridge resistor [B1, p. 214]

Remark: Although the offset is usually expressed in terms of output voltage, it also needs to be considered in terms of magnetic field. A Hall element having a significant offset can still be valid for low magnetic field level applications when it features, at the same time, a very good sensitivity.

### 2.4.3 Temperature dependence

The sensitivity of the Hall probe can be affected by some environmental parameters such as the temperature. The main contribution to this effect originates from the thermal activated behaviour of the carrier concentration. Two cases have to be distinguished:

– The Hall element is made with materials with small band-gap (like InSb for example) and will display strong temperature dependence. It comes from the dependence of the intrinsic carrier's density $n_i$ that increases with the temperature following the activation law

$$n_i \sim T^{3/2} \exp(-E_g / 2kT) . \tag{18}$$

This dependence can be reduced by introducing donor impurities, trying to render the material extrinsic. In that case, the contribution to carrier concentration from thermal activation is negligible and the carrier concentration becomes practically unchanged within a temperature range called the exhaustion range. In this range, the Hall constant $R_H$ is practically unchanged.

– The material that composes the Hall element has a large energy band-gap (like the InAs for example). The semiconductor stays extrinsic up to high temperatures despite low doping.

The temperature coefficient sensitivity and the temperature dependence of the offset voltage are usually expressed by the two following coefficients

$$\gamma_T = \frac{1}{V_H} \frac{\delta V_H}{\delta T} \ [\text{K}^{-1}] \qquad \frac{\delta V_{offset}}{\delta T} \ [\mu\text{V/K}] . \tag{19}$$

As an illustration, Fig. 6 shows the Hall voltage versus temperature of a Hall generator made from InSb biased by two constant currents, 1 mA and 5 mA.

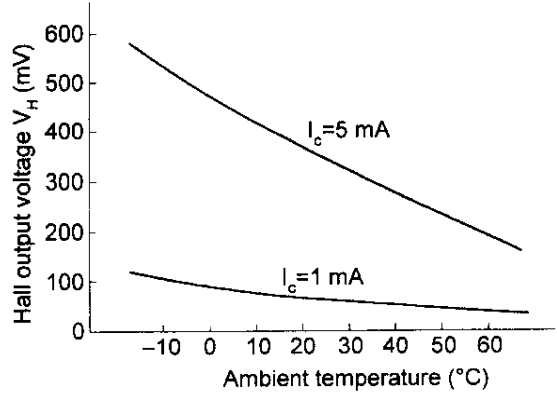

**Fig. 6:** Change of the Hall voltage with respect to temperatures [B2, p.191]

Like the sensitivity, the offset drifts over the temperature but randomly, varying from device to device. Some drifts are caused by piezo-resistive effects. The mechanical stress induced by the thermal expansion of the material leads to changes in the recorded voltages. These variations are not acceptable for some applications when high measurement stability is needed and/or the measurement is long in time. A temperature regulation has to be provided or a correction of the sensitivity must be made.

### 2.4.4  Non-linearity V (B)

The Hall voltage measured in a finite size Hall device is not a linear function of $B$. This non-linearity has various origins. One source is the non-linearity coming from the material property $NL_M$, i.e., caused by the field dependence of $R_H$. Another contribution, $NL_G$, comes from the magnetic field dependence of the geometrical factor $G$. $NL_M$ and $NL_G$ exhibit a quadratic magnetic field dependence but in opposite sign and of the same order of magnitude. The Hall generator is generally designed in such a way that these two effects could compensate each other to a large extent.

$$NL^* = NL_M + NL_G . \qquad (20)$$

The non-linearity is estimated as the deviation of the real voltage at a given biasing current and magnetic field from the best linear fit $V_{Hfit}$ of the measured value. It is usually expressed in terms of percentage by

$$NL = \frac{V_H(I,B) - V_{Hfit}}{V_{Hfit}} *100 \ [\%] . \qquad (21)$$

The non-linearity depends on the Hall generator properties and on the magnetic field interval. High linearity Hall devices will typically display a $NL$ coefficient less than 0.2% at room temperature and for fields below 1 T. This error is determined during the calibration of the Hall probe $V_H$ vs. $B$ and will be corrected (see Section 5.1).

---

[*] NB: In case of vertical Hall sensors built using a CMOS technology (see Section 3.2.1), a third source of non-linearity is the variation with the magnetic field of the effective thickness of the sample $t$ ($B$). The width of the depletion layer between the two sensing contacts changes with $B$, leading to a variation of the sensitivity [8].

*2.4.5 Noise effects*

Noise that appears in the outputs of the signal voltage is one important limitation of the measurement precision. Various sources of inherent noise are generated by the Hall generator itself. The Johnson noise results from thermally induced motion of the electrons in the material. The resulting root mean square of this parasitic voltage $V_{jn}$ is related to the resistance of the device, to the operating temperature and to the bandwidth of the signal by

$$V_{jn} = \sqrt{4 k T R B_w} \qquad (22)$$

where $k$ is the Boltzmann constant ($1.38 \, 10^{-23}$ J K$^{-1}$); $T$ is the absolute temperature in K; $R$ is the resistance in ohms; $B_w$ is the bandwidth in hertz.

To limit this source of noise, the bandwidth must be maintained small, just wide enough to pass the minimum required signal. Johnson noise is limited in the domain of hundreds of nanovolts using a 10 kHz bandwidth and a proper choice of the impedance.

The flicker noise has multiple origins and, in a first approximation, is inversely proportional to the frequency (it is also called $1/f$ noise). Working with low DC frequency signals (for example 1 Hz), this component becomes a significant problem. This noise depends on the material and on the fabrication technique of the Hall element and can therefore be improved. The noise voltage spectral density of a Hall device can be schematically described by the plot in Fig. 7. The corner frequency $f_c$ that limits the two noise regions ranges from 1 to 100 kHz.

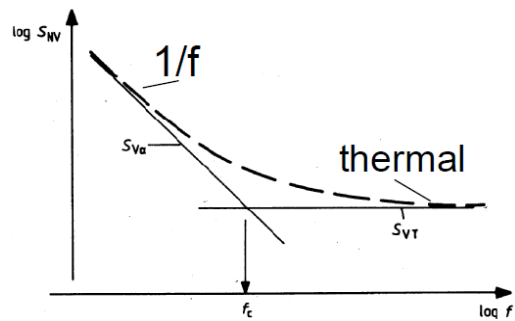

**Fig. 7:** Noise voltage spectral density of a Hall device [B1, p. 215]

## 2.5 'The planar' Hall effect

This galvanomagnetic effect was investigated in 1954 by C. Goldberg and R.E Davis [9] and corresponds to an induced electric field perpendicular to the current in the current–magnetic field plane. The experimental conditions are similar to the 'classical' Hall effect but in that case, a voltage is measured normal to the direction of the current flow *with a magnetic field component in plane made by the current and the voltage measurement direction* (see geometry displayed in Fig. 8). The magnetic field **B** can be decomposed into the following components: the in-plane component $\mathbf{B_p}$ ($B_p \cos \phi$, $B_p \sin \phi$) and the component $B_z$ normal to the plane of the plate, $B_z = B \cos \gamma$. The angles $\phi$ and $\gamma$ correspond respectively to the angles between $\mathbf{B_p}$ and the current direction ($x$ direction) and between **B** and **z** (see Fig. 8). The electric field **E** is expressed by [10, 11]

$$\mathbf{E} = \rho_0 \mathbf{J} - R_H [\mathbf{J} \times \mathbf{B}] + P_H (\mathbf{J} \cdot \mathbf{B}) \mathbf{B} . \qquad (23)$$

With the configuration described in Fig. 8, the transverse coordinate of the electric field $E_y$ is

$$E_y + R_H B \cos \gamma\, J + P_H\, J\, (B_p \cos \varphi)(B_p \sin \varphi) \tag{24}$$

where $R_H$ is the Hall coefficient and $P_H$ a coefficient called the planar coefficient, $\rho_0$ is the magneto-resistivity.

As shown in Eq. (24), the existence of a residual in-plane magnetic field component $\mathbf{B_p}$ generates an additional electric field in the plane perpendicular to $B_p$ and proportional to the square of its intensity. It results in the generation of an additional voltage $U_p$ measured in the transverse direction (y-direction) which is added to the Hall voltage $V_H$

$$V_{output} = V_H + U_P = \frac{R_H\, I\, B \cos \gamma}{t} + \frac{P_H}{2t}\, B_P^2\, \sin(2\varphi)\, I\,. \tag{25}$$

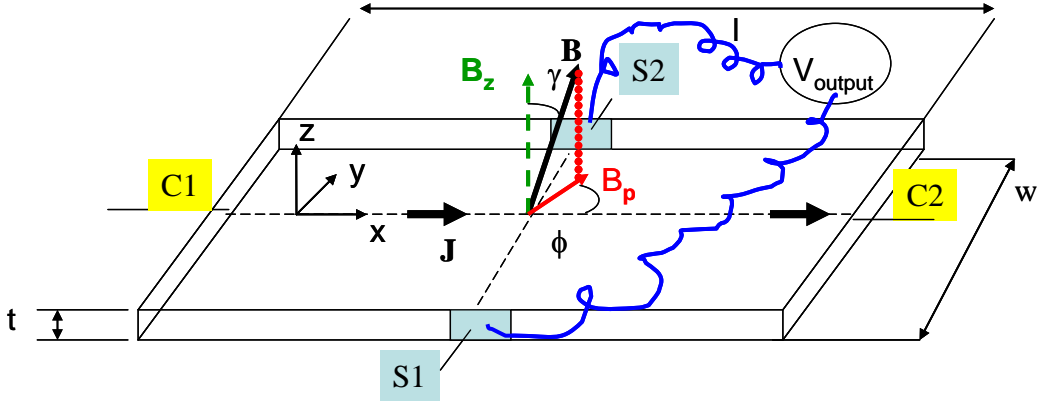

**Fig. 8:** Very long plate exposed to a magnetic induction **B** with an in-plane component $\mathbf{B_p}$

$U_p$ is proportional to $B_p^2$ and follows a $\sin 2\phi$ angular dependence where $\phi$ is the angle between the current and the in-plane component of the magnetic field, as depicted in Fig. 9. This superimposed voltage $U_p$ will spoil the accuracy of the measured value coming from the field component perpendicular to the plane. Planar Hall effect voltage compensation techniques have to be used. As shown by Eq. (25), the planar voltage disappears for $\phi = 0°$ or for multiples of 90°. If the orientation of the in-plane component is known, a rotation of the Hall probe parallel or perpendicular to this direction will cancel this effect. The amplitude of the planar voltage depends on the semiconductor used for sensing. Material with higher conductivity will display a smaller contribution (see amplitudes of $U_p$ in Fig. 9).

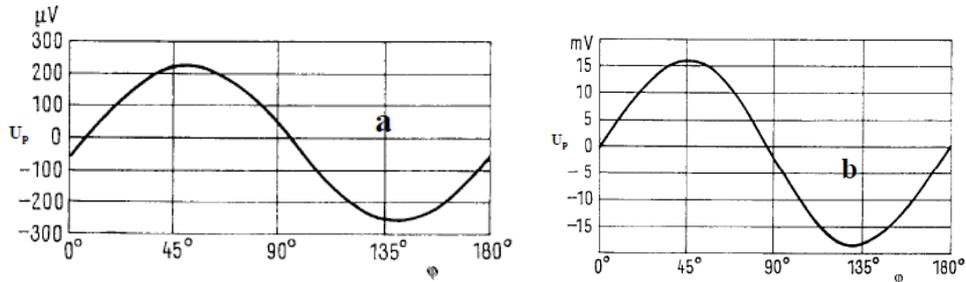

**Fig. 9:** Angular dependence of the planar voltage $U_p$ for an InSb Hall probe for two different conductivities: (a) n-doped InSb $\sigma = 750\ \Omega^{-1}\text{cm}^{-1}$, (b) intrinsic InSb, $\sigma = 200\ \Omega^{-1}\text{cm}^{-1}$ (from Ref. [7])

Compensation techniques were proposed to cancel the planar Hall effect. The use of two identical Hall probes on the top of each other, but the lower turned by 90° w.r.t. the upper one, was suggested by Turck [10] and described in Ref. [12]. The direction of current of probe 1 and probe 2 are perpendicular and make an angle ɸ$_1$ and ɸ$_2$ = π/2 - ɸ$_1$ with the planar component of magnetic field.

In that case, the compensation occurs because sin(2 ɸ$_1$) + sin(2ɸ$_2$) = 0. The limitations of this arrangement are listed in Ref. [13]. In particular the Hall probe material and the geometry must be identical. Moreover the centre of the probe must be located on the same line perpendicular to the active part and the angle of rotation has to be exactly 90°. An arrangement using a single sensor, where the voltage and current connection are interchanged on the fly, is described in Ref. [14].

## 2.6 Quantum Hall effect (field measurements at cryogenic temperature)

The (integer) quantum Hall effect is observed in two-dimensional electron systems subjected to low temperatures and strong magnetic fields. The measured Hall conductivity σ shows in that case quantized values. This phenomenon was discovered in 1980 at the Grenoble High Magnetic Field Laboratory in France by Klaus von Klitzing [15]. Studying the Hall conductance of a two-dimensional electron gas at very low temperature, he observed that the dependence of the Hall voltage as a function of the magnetic field displays a staircase sequence of wide plateaux instead of the expected monotonic increase (see Fig. 10). These steps occur at precise values of conductance equal to $ne2/h$, where $n$ is the integer that characterizes each plateau, $e$ the electron charge and $h$ the Plank constant. The integer quantum Hall effect can be explained by the quantization of the electron orbits in two dimensions. These orbits correspond to Landau levels which have the discrete energy values

$$E_n = h/2\pi \, \omega_c (n+1/2). \qquad (26)$$

$\omega_c = eB/m$ is the cyclotron frequency.

The application of strong magnetic fields leads to a degeneracy of each Landau level and the electrons have many possibilities for a state occupation. The change of electronic levels gives rise to 'quantum oscillations' called the Shubnikov–de Haas oscillations (see Fig. 10). Such oscillations limit the accuracy of the Hall voltages measurements. The following plot displays the deviation $\Delta V$ of the Hall voltage from a straight line as a function of the applied field resulting from these oscillations. The perturbation can reach up to a level of one per cent at high fields, depending on the semiconductor used for the sensing. A precise calibration at cryogenic temperatures is necessary for a low temperature and high accuracy of magnetic field determination.

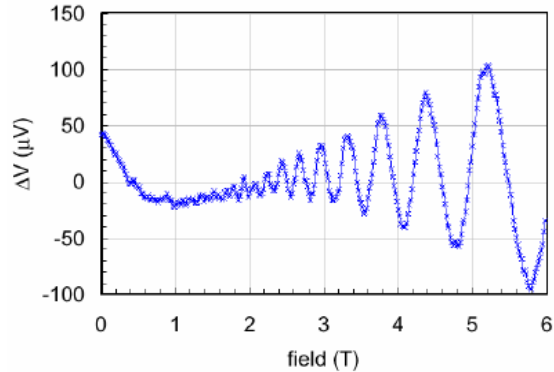

**Fig. 10:** Shubnikov–de Haas effect in a Hall generator at 4.2 K. The plot displays the deviation ΔV of the Hall voltage from a straight line as a function of the applied field (reprinted from Ref. [16])

# 3 Single-axis Hall probe

Hall probes, i.e., Hall generators mounted on a sample holder, are available in a variety of shapes and active area dimensions.

## 3.1 'Plate-like' Hall probes

The use of a sensor with a 'plate-like' geometry is optimal for measurements of homogenous magnetic field (at the level of the active part), provided that the sensor can be positioned correctly. The families, transverse and axial, are defined by the direction of the flux lines through the probes (see Figs. 11 and 12). The Hall element is mounted at the end of a long flat stylus. The transverse type is designed for the measurement of the main magnetic field component perpendicular to the chip surface and is sufficient for many applications. This Hall plate has a quite simple structure and can be adapted to almost any kind of semiconductor technology without special requirements or additional steps. For this reason the devices are generally very cheap to manufacture.

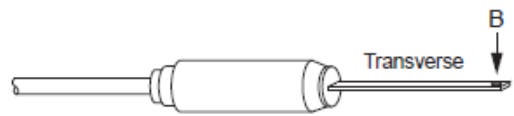

**Fig. 11**: Transverse Hall probe (picture from Ref. [6])

The axial type is generally cylindrical, the Hall plate being mounted perpendicular to the axis of the cylinder. It is designed to measure fields parallel to the axis (Fig. 12).

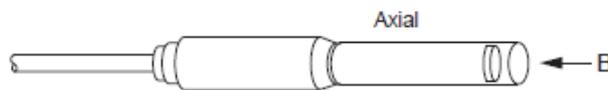

**Fig. 12**: Axial Hall probe (picture from Ref.. [6])

At the beginning of the 1980s other geometries were proposed, based on the following general criteria established for good Hall devices [17]:

– In the active part, the current density and the magnetic field have to be as much as possible close to the orthogonality.

– The shape of the active region has to facilitate the measurement of the Hall voltage $V_H$.

– Offset voltage at zero fields has to be minimized, i.e., along the sensing contact (S1S2), the electrical field **E** and the Hall field $\mathbf{E_H}$ have to be as much as possible close to orthogonality.

## 3.2 Non 'plate-like' devices

The conventional shape of the Hall generator mentioned in Section 2.3.1 (rectangular, cross, bridge, etc.) is sometimes not adapted in the case of a non-homogeneous magnetic field or a particular shape of the field lines. A second class of Hall generators, called non-plate-like, features a tri-dimensional structure. They are used to fabricate sensors capable of measuring simultaneously two or three components of the magnetic field. The devices are structured in order to fulfil the three criteria presented above. Reference [17] is a survey of this second class of Hall generator.

## 3.2.1 Vertical Hall Device (VHD)

The vertical Hall device (VHD) was devised more than 20 years ago (see Refs. [18, 19]) to respond to a magnetic field parallel with the chip surface. The device is called 'vertical' as the region that generates the Hall effect is perpendicular to the chip plane. It can be obtained from a transverse Hall plate by the transformation schematized in Fig. 13:

–  turning the horizontal Hall plate into a vertical position;
–  dividing one biasing contact into two parts and arranging all the contacts in line on top of the plane.

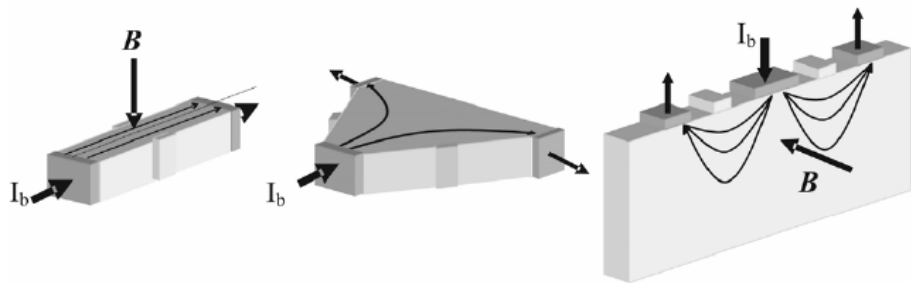

**Fig. 13:** Transformation of a plate-like Hall probe into a vertical one (from Ref. [20])

The advantage of this geometry is that all the electrical contacts are on one side of the device so it is compatible with standard one-side silicon technology. The use of vertical sensors allows one to fabricate the devices by evaporation, sputtering, and the pattern is structured using photolithography. With integrated circuit technology, not only thin devices with high sensitivity but also deeper structures can be obtained by implantation and thermal diffusion of impurity ions. Vertical Hall sensors with high current sensitivity $S_I$ = 400 V/TA have been produced by Landis and Gyr (Switzerland). A patented technology called 'Vertical Hall Technology' based on n-type silicon wafers with low doping concentration like $2 \cdot 10^{14}$ cm$^{-3}$, was employed. The technology and the device are described in Ref. [20].

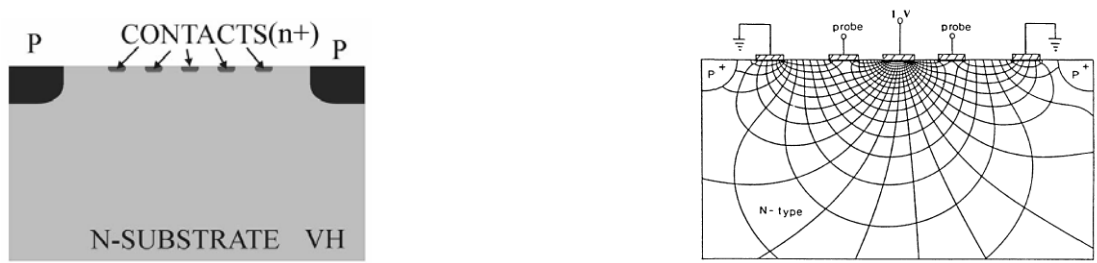

**Fig. 14:** Left: Vertical 'bottomless' Hall sensor made by the VH technology (from Ref. [20]). Right: Principle of a vertical Hall sensor. The current flow and the equipotential lines are shown. The Hall voltage is measured between the two probe contacts (from Ref. [20])

Figure 14 (left) displays the cross section of a vertical Hall sensor built with this process. Figure 14 (right) describes the principle layout of a vertical Hall sensor with **B** perpendicular to the plane. The active zone belongs to the n-type substrate material, limited by a p-well ring. In the version presented in Fig. 14, the VHD has five contact terminals made by n+ implantation and arranged in a line on top of the active diffusion region. The current is injected through the central contact to the two outside ones connected to ground. The Hall voltage is sensed between the probe contacts. The output voltage $V_H$ is given by the integral of an equipotential line between the two sense contacts. However, production costs turn out to be high due to the special process of obtaining low doped and deep structures. Moreover, the sensor is open downward and allows a deep current flow (e.g. 30 μm). Other technologies like bipolar [21] or CMOS technology [22, 23] have been used to manufacture vertical devices with the key advantage of allowing the co-integration of the electronics on the same chip.

From Eqs. (10), (16) and (17), it turns out that current sensitivity $S_I$ is inversely proportional to the product of the thickness $t$ and the doping concentration $n_d$:

$$S_I \propto \frac{1}{t\, n_d}. \tag{27}$$

To increase the sensitivity, the first obstacle to overcome is therefore to limit the depth of the biasing current flow. Using the CMOS technology with an additional deep-n well, the device could be isolated from the substrate by the layer and the current will therefore be concentrated near the surface (between 0 μm and 3 μm). The second difficulty is to use a low doping n-layer. A patented method [24] involving a different structure of the implementation mask reduced the level of doping by almost one order of magnitude near the surface. A complete description of the CMOS process [22 and B3] and the performance of vertical devices fabricated with the CMOS technology can be found in Ref. [25]. Variations of the sensor layout have also been developed. A design with four contacts allowing the four-fold symmetry advantages for the offset compensation was proposed [8]. An improved version features a design with six contacts with two outside contacts short-circuited to minimize the offset voltage. Figure 15 displays the three design structures of a vertical Hall sensor, realized using the integrated circuit technology: five terminals (centre), four terminals (left), six terminals (right).

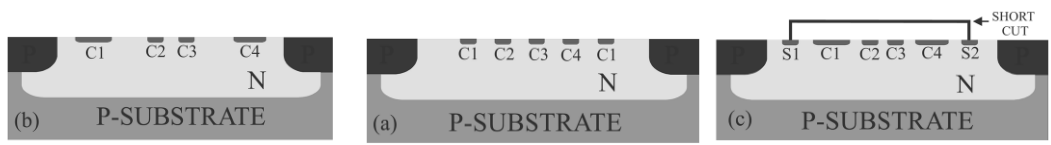

**Fig. 15:** Structure of a vertical sensor sensitive to a magnetic field parallel to the surface plane built using a derived CMOS technology. Three variations are presented: the design with five terminals (a), four terminals (b), six terminals (c) [from Ref. B1 p. 263]

This technology allows the manufacture of large quantities of Hall devices with very small dimensions (micro-sensors) and for a low price. Sensors fabricated with the integrated circuit technology have highly reproducible geometry and physical properties. Figure 16 is an example of a vertical device.

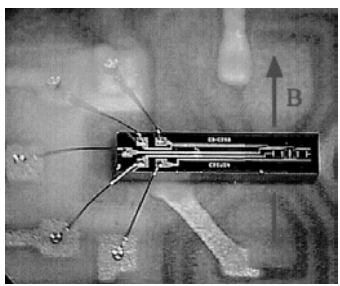

**Fig. 16:** Vertical device sensitive to a magnetic field parallel to the surface of the chip from Ref. [17]

### 3.2.2 Cylindrical devices

A cylindrical Hall device features a magnetic field sensitive volume in the form of a half-cylinder imbedded in the sensor chip. It results formally from the conformal deformation of a vertical device (see Fig. 17). This shape of sensor was devised to be sensitive to circular magnetic field form. A typical application is the current sensor.

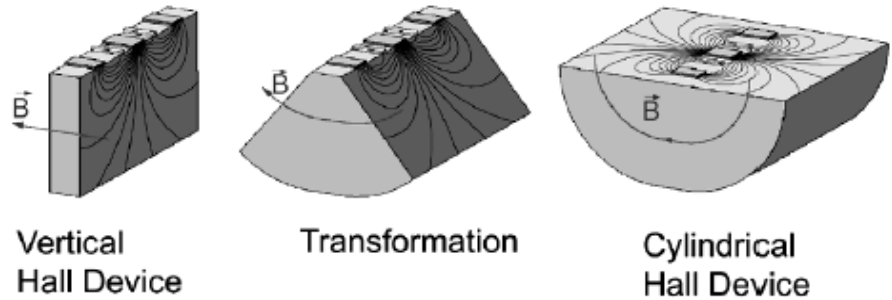

**Fig. 17:** Cylindrical device (right) obtained from a vertical Hall device (left). The sensitive volume of a cylindrical device is adapted to the measurement of a circular magnetic field [from Ref. B1 p. 265]

### 3.2.3 'Planar Hall' device with open bottom

This Hall generator has the aspect of the Hall plate but features an open bottom structure, i.e., a structure not isolated at all from the bulk. The active zone is deep and can be compatible with a vertical Hall device to form a three-axis sensor. Figure 18 illustrates the structure of the so-called 'planar Hall' device made with the technology for vertical devices (see Section 3.2.1). The design has a cross shape to make the sensor invariant for a switch between the biasing and sensing contacts.

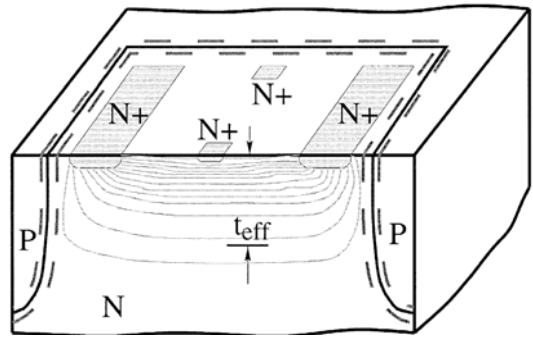

**Fig. 18:** Planar Hall device made with the technology of vertical devices. The effective thickness $t_{eff}$ defines the depth down to which the current is forced to flow [26]

As for the vertical device, the active zone is limited laterally by the P-well ring but not vertically. To increase the sensitivity, the spread of the current into the bulk has to be limited to an effective thickness $t_{eff}$ just below the chip surface. The reduction of the size of the sensor together with an additional surface doping, results in a well-defined conductive zone below the surface. The sensor can reach a current-related sensitivity of 0.33 V/mA T [26].

# 4 Two- and three-axis Hall effect devices

## 4.1 Two-axis device

A two-axis Hall device that measures simultaneously the two in-plane components $B_x$ and $B_y$ can be obtained by associating two vertical bottomless generators orthogonally. The combination to generate a two-axis device is displayed in Fig. 19.

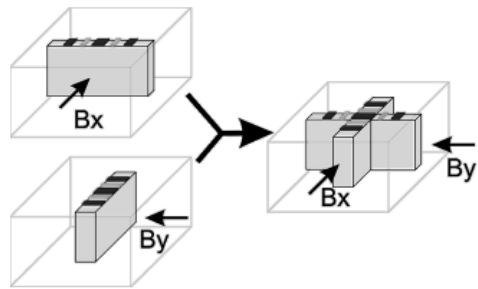

**Fig. 19:** Fabrication of a two-axis Hall device [from Ref. B1 p. 262]

Combining such a magnetic sensor with a permanent magnet, one can fabricate an accurate contactless angular position sensor [17].

## 4.2 Three-axis sensor design

### 4.2.1 Conventional geometry

A combination of three Hall magnetic generators oriented in the three main directions and supported by three orthogonal plates was and remains a classical way to measure simultaneously the three components of the magnetic field. This arrangement (see Fig. 20) has several disadvantages:

– The spatial resolution is limited by the distance between the single elements.
– The precision in the orthogonality of the three sensors will depend on the assembly procedure.
– The measurement is not 'point like' as the flux density is not measured at the same spot.
– The number of contact leads is high.

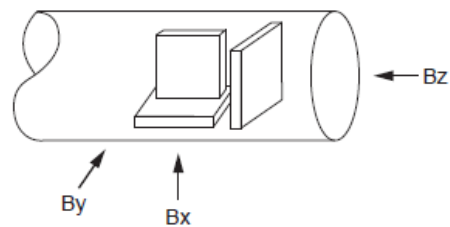

**Fig. 20:** Three-axis Hall probe built by associating three single-axis Hall probes (from Ref. [6])

The main advantage with this construction is that the accuracy of a single Hall probe can be practically kept in the three directions.

*4.2.2 Geometry using vertical Hall devices, micro Hall sensors*

With the development of vertical Hall devices it is possible to fabricate an integrated structure by merging two or three modules. The planar technology is based on the principle that the magnetic field comes from the decomposition of two perpendicular in-plane components and a third orthogonal to it. The resulting 'new generation' of Hall sensors features a reduced active measurement volume and an improved orthogonality of the axes of sensitivity. A tri-axial chip with a square shape, built using a small single volume of a semiconductor crystal, was devised in 1999 [20]. The working principle consists in the association of the vertical and plate-like structure. For the measurement of the in-plane components called $B_x$ and $B_y$, the geometry is similar of that a vertical device. For the component perpendicular to the sensor plane, the geometry is derived from that of a Hall plate but with an open bottom. As displayed in Fig. 21, the structure consists of a square shape silicon block with four current contacts in the corners and four sense contacts between them.

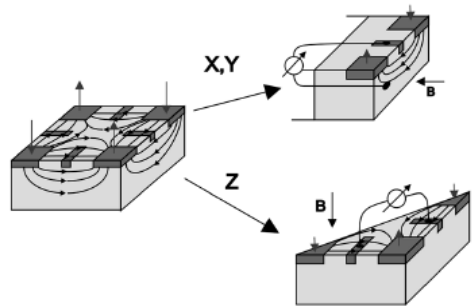

**Fig. 21:** Basic structure of a tri-axial Hall device. From the recorded voltages $V_x$, $V_y$, $V_z$, the three components of the magnetic field are obtained. The bias contacts are located on the corners of the square and the sense contacts are between them. The arrows indicate the flow direction of the currents [from Ref. B1, p. 266]

Similarly, three-axis Hall chips are made using two different types of devices as shown in Fig. 22. A VHD is merged with a planar Hall device whose geometry is described in Section 3.2.3.

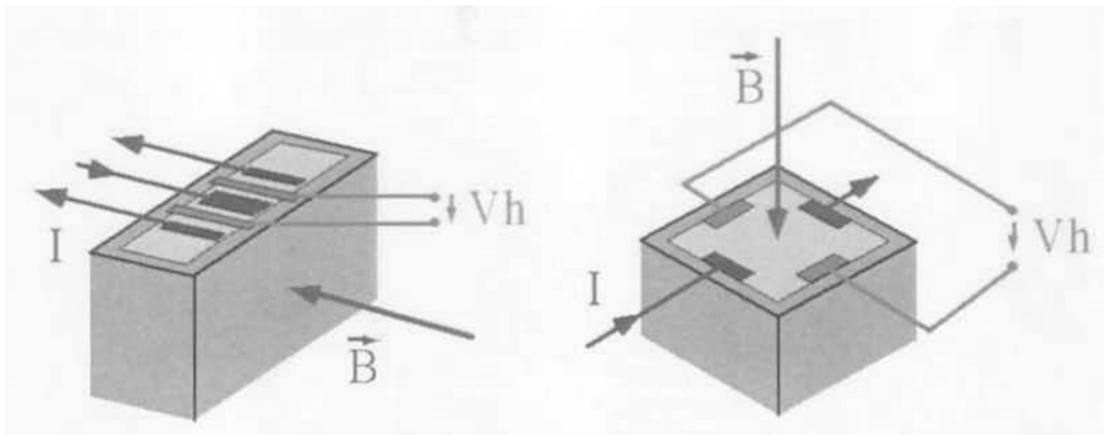

**Fig. 22:** Vertical Hall element (left) and a Hall planar device, i.e., a conventional Hall generator with an active part open to the bottom (right). The two types of element implemented in a chip will measure the three components of the magnetic field. The pictures are reprinted from Ref. [20]

The possibility of co-integrating a Hall generator and smart electronics on a single chip allows the fabrication of integrated Hall sensors, having dimensions typically between 10 μm and 100 μm. A picture of such a Hall element fabricated with the CMOS technology is shown in Fig. 23. Hall cross shape sensors, featuring an active area as small as 0.1 μm square is even achieved. The performance, the technologies and the applications of the micro Hall devices are reviewed in Refs. [27, B5].

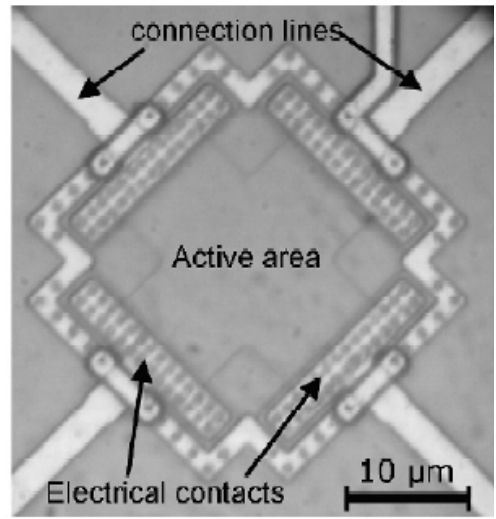

**Fig 23:** CMOS Hall sensor (from Ref. [B3], p. 4)

The small size of these micro-sensors facilitates the integration of a large number of Hall elements in the same chip surface. Therefore, another solution for multi-axis sensors consists in spacing the two types of device for each axis of sensitivity. To assure a measurement for each component in the same spot, the sub-sensors have to be arranged in pairs at equal distance from the measurement centre. For example, the structure shown in Fig. 24 is composed of eight Hall sensors on silicon [20]. The components that are parallel to the sensor surface are measured by two pairs of vertical Hall devices located on the opposite sides of the sensor system. The perpendicular component is measured by four horizontal devices located in the corner of the system.

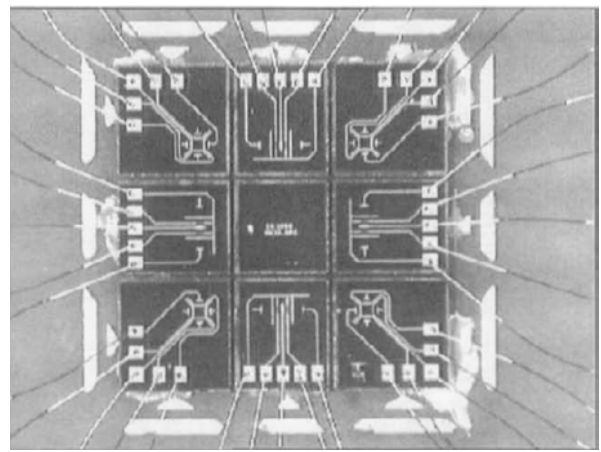

**Fig. 24:** Three-axis sensor made of eight sub-sensors The two sensor types are arranged to share one common active volume, formed by the eight sub-sensors. The dimensions are of the active part are $2 \times 2 \times 0.2$ mm$^3$ (from Ref. [20])

## 4.3 Example of a three-axis integrated circuit Hall sensor

The three-axis IC Hall sensor produced by the company SENIS is presented here as an illustration, see Refs. [28, 29]. The sensing part is composed of one single horizontal Hall element at the centre which senses the component perpendicular to the magnetic probe. Two pairs of vertical devices aim at measuring each in-plane component of a magnetic field (see Fig. 25). The devices are fabricated using the N-well CMOS technology. The dimensions of the sensing part are about 150 µm × 150 µm. The sensitive volume is 0.15 mm × 0.15 mm × 0.01 mm and the mutual orthogonality of the three axes is below 0.5°. The sensing part is integrated into a complex electronic circuitry which supplies the current, provides the amplification and the cancellation of parasitic signals like the offset and the $1/f$ noise (see Section 4.4.1). A temperature sensor is also integrated in the chip. The dimensions of the chip are 6400 µm × 6400 µm × 550 µm.

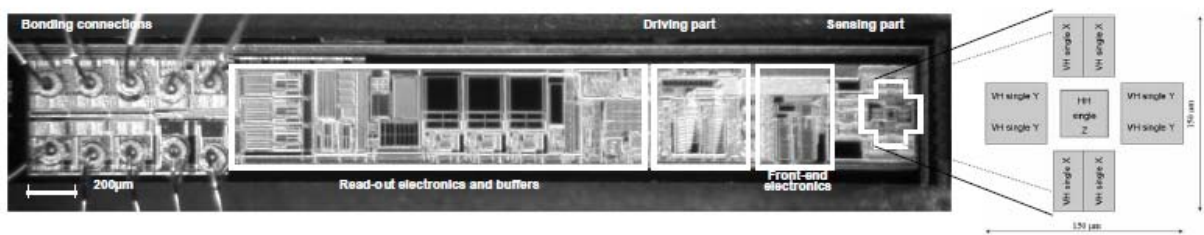

**Fig. 25:** Photo of a three-axis Hall probe chip produced by the company SENIS (from Ref. 29])

The Hall sensor features a spatial resolution of 0.15 mm, a negligible offset voltage and planar Hall effect. The IC Hall sensor has equipped a three-axis teslameter, commercially available with a standard accuracy of 0.1% in the magnetic field range between 20 mT and 2 T.

The performance of this teslameter is summarized in Table 2.

**Table 2:** Specifications of a commercial teslameter from SENIS [29]

| Full scale (nominal) | 2 T |
|---|---|
| Sensitivity to D.C magnetic field | 5 V/T |
| Tolerances of sensitivity ($B$ = 1T, D.C) | 0.1% |
| Temperature coefficient of sensitivity | <100 ppm/°C |
| Residual non-linearity (up to 2 T) | <0.05% |
| Planar Hall effect: $V_{plan}/V_{vert}$ (1 T) | <0.01% |
| Offset (B = 0 T) | <± 1 mV (± 0.2 mT) |
| Long term instability of sensitivity | < 1% over 10 years |

This new generation of Hall magnetometers equipped with IC Hall sensors is now commercially available, offering the possibility to measure all three components of the magnetic field even in highly inhomogeneous fields due to strong reduction of the active volume with respect to the conventional Hall probes. The technological developments underlying each of the components of this new type of magnetometer system are reviewed in Ref. [30].

## 4.4 Improvements in the three-axis IC sensor characteristics

The most important issues are offset, noise, temperature dependence, and planar Hall effect.

*4.4.1    Offset reduction*

Hall sensors made with CMOS technology suffer from large offsets which make them less appropriate in high accuracy applications. As mentioned in Section 2.4.2, the offset has many origins: material inhomogeneities, material construction, mechanical stresses, and temperature variations. These resulting offset fluctuations can hardly be compensated even with frequent calibration. Offsets at the level of few mT are typically observed and realistic signal levels of a few µV used in some applications become then completely masked. The orthogonal coupling of the sensors is the traditional static offset cancellation (reduction) method. The coupling is done by the direct connection of two sensors, each having a biasing mode shifted by 90° [31]. Figure 26 describes the equivalent circuit using the electrical analogy of the bridge resistor. The offset cancellation is provided by the matching properties of the two sensors. For a successful compensation, the increment of resistance $\Delta R$ that appears in one leg of the bridge has to be equivalent in the two sensors.

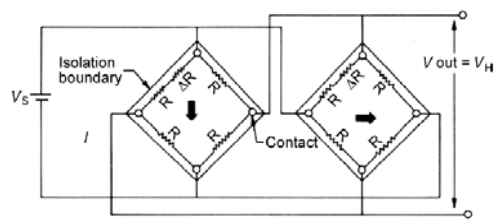

**Fig. 26:** Pairing of two matching sensors. They are connected electrically
in parallel but with orthogonal current directions [31]

This compensation gives approximately a tenfold improvement over the uncompensated Hall probes, cancelling in particular the effect of mechanical stresses. This technique is limited by the mismatching between the Hall devices. The drift of the offset could not be strictly equivalent for the two Hall generators and the resulting compensation will only be partial. A step further is the use of the so-called 'spinning current Hall plates' developed at Delft University in 1992 [32]. As shown in Fig. 27, the Hall generator is symmetrical and the current terminals are periodically swapped with terminal voltages. In the first cycle, the biasing current flows between points 1 and 2. The output voltage between points 3 and 4 is $V_{out1} = V_H + V_{offs}$. In the second cycle, the connections are commutated and an output voltage $V_{out2} = V_H - V_{offs}$ is recorded between the points 1 and 2. The difference between the mean values $V_{out1}$ and $V_{out2}$ is equivalent to $V_H$. The offset reduction is achieved by the fact that Hall voltage does not depend on the current direction, but the offset voltage does. Using four terminal devices, the offset can be reduced to 0.5 mT. Spinning-current Hall plates with eight contacts were also designed. With the association of dynamic offset cancellation techniques like auto-zeroing and chopping [32], the offset can be reduced to values down to 1 µT [33].

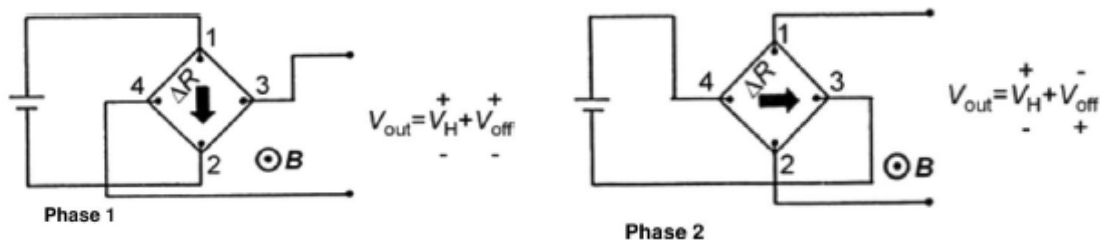

**Fig. 27:** Spinning current technique for offset compensation: (a) phase 1, current is flowing from 1 to 2 and the voltage is sensed between 3 and 4. (b) phase 2, the terminals are swapped (from Ref. [29])

*4.4.2   Compensation of the temperature dependence of the sensitivity*

The simplest way to compensate the variation of the sensitivity with temperature is the use of a resistor connected in parallel with the input terminal of a Hall device. The key point is that at constant bias current, the change in the input resistance of the Hall device will modify proportionally the output voltage. The input resistance, defined as the resistance measured between the current leads, will be used as a temperature indicator. As shown in Fig. 28, the current $I$ is divided into two parts: $I_{in}$ flows in the Hall element and $I_R$ in the resistor. The change of the input resistance caused by the temperature variation will affect the current redistribution among $I_{in}$ and $I_R$. Thus by adjusting the bias current $I_{in}$ the change caused by the temperature effect is compensated.

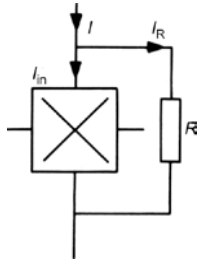

**Fig. 28**: Simplest compensation scheme for temperature effect compensation on sensitivity (from Ref. [B1], p. 281). The square with the cross is the symbol of the Hall device (the output is the result of the multiplication of the biasing current and of the magnetic field)

More sophisticated circuits can also be built with an operating amplifier and a differential magneto-resistor scheme [B5].

*4.4.3   Reduction of the 1/f noise*

The 'spinning current' technique (Section 4.4.1) also reduces the $1/f$ noise. The $1/f$ noise can be seen as a slowly fluctuating offset voltage, appearing alternately in the output voltage terminals when the bias current is rotated back and forth. Provided that the switching frequency is higher than the corner frequency $f_c$ (see Section 2.4.5), the voltage noise contribution, together with the offset, can be separated from the useful signal and filtered. The spinning frequency is high, several hundred kHz.

*4.4.4   Hall planar effect reduction in three-axis Hall sensors*

Evaluation of the contribution of the planar Hall effect is of a crucial importance if one uses three-axis Hall sensors. For vertical devices made with silicon, Schott found a theoretical contribution of 12%. This estimation was not confirmed by the measurements, indicating, on the contrary, a lower contribution down to 1%. This strong reduction was attributed to the particular spatial geometry of the vertical Hall device and its resulting current paths in the vertical structure [34]. Unlike the case of the plate shape, only a short section of the current filaments are distorted by the planar flux density, thus reducing the planar Hall effect. Current spinning also cancels the planar Hall voltages. We refer again to the corresponding magneto-resistance model presented in Fig. 29. Let us take the case of a Hall element with a cross geometry. The current $I$ is sent from one contact to its opposite one and the output signal is measured between the two other contacts. The model is made by four resistors having, for simplification, an equal value at zero-field. A planar magnetic flux density $\mathbf{B_p}$ acting on the diagonal direction will unbalance the resistor bridge by changing the resistance values. To a first approximation, the magnetic field will indeed change the value of the equivalent resistor only if a field component perpendicular to the current flow through the resistor is present. This unbalance will create a supplementary voltage $U_p$ proportional to $B_p^2$ (as recalled in Section 2.5), appearing between two of the four terminals.

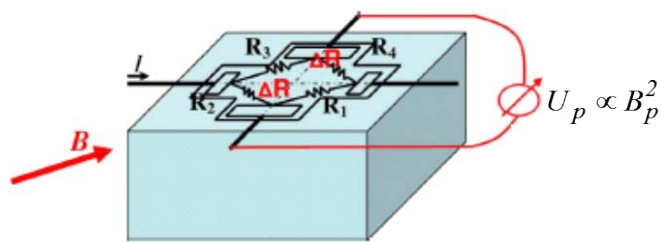

**Fig. 29:** Contribution of the Hall planar effect to the output voltage on a Hall element electrically modelled by a resistor bridge (picture from Ref. [29])

The similarities with the situation described in Fig. 5 for the offset led to the investigation of the impact of the 'spinning current' method on the planar Hall effect contribution. The result for a horizontal Hall plate is presented in Fig. 30. The amplitude of the planar Hall voltage $V_{PH}$ is determined with and without spinning current effect compensation, for fields up to 2 T. The compensation works successfully with reduction of the ratio planar/conventional Hall voltage from 1.3% down to 0.02%.

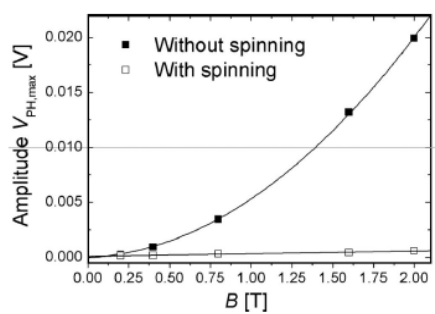

**Fig. 30:** Measured planar Hall voltage without and with the application of the spinning current technique [29]

### 4.5 Three-axis Hall sensors: limitations and the next challenges [35]

- For the Hall probes built in the conventional way (Section 4.2.1), one main improvement concerns the physical size. The merging of three single-axis sensors sets a lower limit on the size of around a few mm³. This not only limits the gap size that one can map but also the capacity of the sensors to be used in highly inhomogeneous fields. The distance between the three sensors leads to a different measurement of the field and therefore, the measurement is not a 'three-axis' consistent point measurement. The accuracy of the orthogonality is another critical issue. The orthogonality in a traditional three-sensor assembly is not easy to achieve with sensors that have such a small surface area. A tilt error of 1 μm on a sensor of 1 mm diameter may already be detectable.

- In the three-axis IC Hall sensor, *sensitivity and accuracy are* the major concern. The size constraint limitation combined with the less efficient vertical Hall probe geometry and the material choice for the integrated sensor (Si), renders this type of sensor difficult to use for high-accuracy field measurements (e.g., at 0.01% of the measurement range). Multi-axis IC sensors typically feature an accuracy between 0.1% and 1%, one or two orders of magnitude less than conventional single-axis probes. A standard accuracy of 0.01% (of the measurement range) has been difficult to achieve up to now. Noise coming from the circuitry, drift of the offset due to temperature effects and difficulty in the calibration to remove the effects caused by the mutual interaction of the three axes are the major limitations.

- Cryogenic applications also need special attention. Some integrated sensors 'freeze out' at low temperatures, and are therefore not suitable for cryogenic applications. The oscillations coming from the quantum Hall effect (Section 2.6) have also to be considered. Cryogenic multi-axis sensors have therefore, up to now, been built in the traditional way (three single-axis sensors), with carefully chosen bulk semiconductor material, and with severe constraints on the mechanical stability to be able to survive the temperature cycling.

- Finally, the calibration of three-axis devices remains a complex and long process (see next section). Up to now, the end-user generally does not have the equipment, the time and the knowledge required and will rely upon the manufacturer. The calibration of the three major axes at high magnetic fields is not easy since the small gap of these magnets makes it difficult to insert the probe. In addition, a good mechanical reference surface is required in order to have a reproducible position of the probe in all three orientations. To improve the sensitivity down to 0.01%, the cancellation of the planar Hall effect and of the high order of perturbations is required. The method necessitates being able to accurately orient the calibration field, not only along the major axes but also at many angles in between. Furthermore, reliable calibration software that reconstructs coherent calibration parameters — usually nonlinear — based on all these measurements is needed. A considerable effort has to be made to simplify the calibration process using commercially available tools.

## 5    Calibration of the Hall probes

### 5.1    Single-axis Hall probe calibration

The calibration is necessary to determine the real output signal of the Hall probe. As recalled in Section 2.4.4, the Hall voltage is not a linear function of *B* and is a function of the temperature. These functional dependences of the Hall voltage have to be known before starting the measurements. The calibration is the test during which known values of the magnetic field are applied to the Hall probe and the corresponding output reading is recorded. This procedure takes place in all field ranges of interest and is repeated quite often to avoid problems with offset drifts. A successful calibration method consists in using as a reference the magnetic field produced by an NMR teslameter *B-NMR*. Typically, one needs:

– a Hall probe current source with a high stability;
– a high-accuracy voltmeter to measure the Hall voltage;
– an NMR teslameter with probe(s) to measure the applied magnetic field;
– a calibration magnet with its power supply. The required homogeneity of the magnetic field $\Delta B/B$ is $10^{-4}$, $10^{-5}$ per mm for the locking of the NMR frequency;
– a temperature-stabilized room.

The Hall and NMR probes are placed in the gap of a magnet whose field magnitude and polarity can be changed while the readings of $V_H$ and *B-NMR* are recorded. At the Paul Scherrer Institut, the calibration field range is from $\pm$ 0.1 T to $\pm$ 2 T. The low end of 0.1 T is our measurement field limit of the NMR teslameter (Metrolab PT 2025 with probes 2-5). The high end at 2 T is due to the field non-homogeneity of the calibration magnet (Bruker dipole B-E 30hf), which at this level, is too high for the teslameter to obtain the NMR signal. The Hall and NMR probes are placed in the centre of the pole gap, each on one side of the magnet mid-plane and at equal distance from the magnet poles (Fig. 31). In this way, both probes experience the same field value. Any possible variation of the magnetic field from mid-plane to pole does not therefore affect the calibration results. The Hall probe needs to be oriented perpendicular to the field, otherwise the measured value will correspond to $B \cos(\gamma)$, where $\gamma$ is the angle between the field and the axis normal to the probe. An error of 1 degree reduces the $V_H$ reading by $2.10^{-4}$. The NMR is insensitive to small-angle errors.

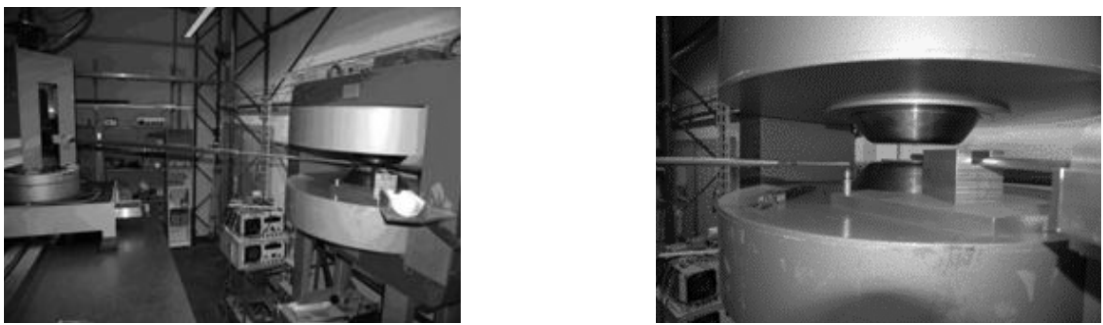

**Fig. 31:** Left: Hall probe inside a 2 T calibration magnet. The magnet has a gap of 38 mm and a field homogeneity $0.6 \times 10^{-5}$ in the centre. Right: Measurement of the Hall voltage $V_H$ for a magnetic field read by a NMR probe *B-NMR*

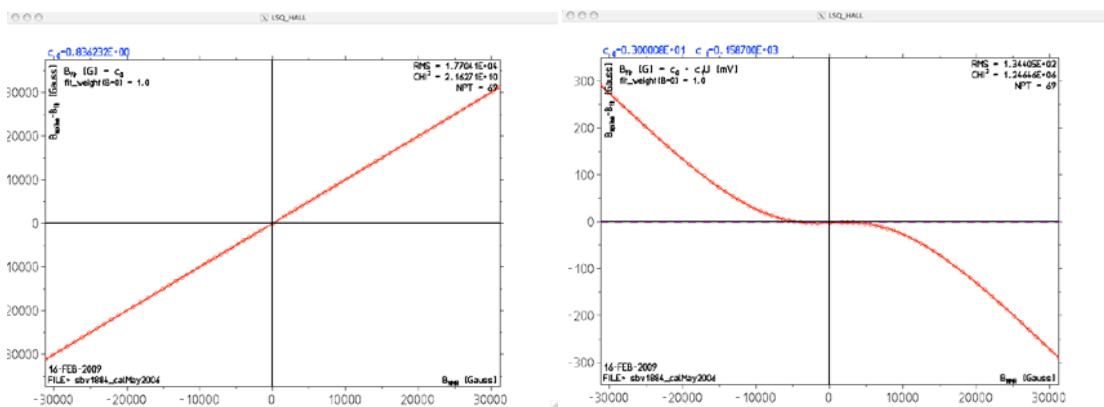

**Fig. 32:** Left: Hall voltage recorded versus magnetic field read by the NMR. Right: Deviation from the linearity $V_H = f(B)$ in the case of the Siemens SVB601 Hall probe. The non-linearity is less than 0.5% up to 2 T

During the calibration (and afterwards during the magnetic measurements) the environment air temperature should be stabilized and kept constant. While the Hall voltage is not only directly proportional to the magnetic field but also to the Hall probe current, a high-precision constant current source for the Hall probe is required (<100 ppm). In order to improve the accuracy of the $V_H$ reading, a method with averaging of multiple measurements should be used. An integration time of 20 ms in the digital voltmeter (DVM) will filter the 50 Hz noise that is present in every electrical device so it should be chosen if possible. Figure 32 shows an example of deviation from the linearity $V_H = f(B)$ in the case of the single-axis Hall probe Siemens SVB601, used at the Paul Scherrer Institut.

Several methods are used to describe the non-linearity of the field versus Hall voltage. The simplest and easiest to use is

$$B = c_0 + c_1 V_H + c_2 V_H^2 + \ldots c_n V_H^n, \qquad (28)$$

Where the coefficients $c_i$ are calculated with the least-squares fit of the tabulated measured $V_H$-*Hall*/*B-NMR* data.

The highest order term *n* is typically between 10 and 15. The higher the order of polynomial, the better the approximation. Above a certain order the improvement is, however, not significant. The highest order term can be determined when the reduced Chi-square $\chi^2$ of the fit is near 1.

$$\chi 2 / ((N - n) B^2_{err}) \approx 1 \quad . \tag{29}$$

Here *N* is the number of measured pairs ($V_H$, *B-NMR*) and $B_{err}$ the field measurement error.

## 5.2 Calibration of a multi-axis Hall sensor

In the case of a three-axis sensor, the difficulty of performing a precise calibration is increased because of the presence of the mutual coupling between axes. The classical operation consisting in the calibration of the probe along the three main axes is not sufficient to reach a level of precision better than 1% because contributions like the planar Hall effect, which are not symmetric around the axis perpendicular to the Hall probe plane, are zero on these main axes. A dedicated patented calibration device was developed at CERN [36, 37] in order to estimate and cancel out the high-order terms of the Hall voltage. In the proposed calibration technique, the sensor makes a three-dimensional scan, being turned in a constant and homogeneous magnetic field in steps of a fraction of π like π/12 for example. The operation is repeated for several field strengths and temperatures. Then the output voltage is expressed in orthogonal functions, the spherical harmonics for the spatial dependence, and the Tchebychev polynomials for the magnetic field and temperature dependence. The set of equations (e.g., set of voltages) obtained for the various angular steps and different magnetic fields is solved and reduced to the different components of the decomposition. The correspondence between the main spherical harmonics $Y_{lm}$ and the components of the output voltage are presented in Table 3. This calibration process aims at correcting the errors coming from the non-linearity, the non-orthogonality of the sensors, the temperature sensitivity, and the planar Hall effects. The three spatial components of **B** can be reconstructed with a precision of about 0.01%.

**Table 3**: Coefficients of the output voltage components resulting from decomposition in spherical harmonics. The biasing current flows in the *x* direction and a classical 'Hall voltage' is measured in the *y* direction

| Coefficient $Y_{lm}$ | Significance |
|---|---|
| $Y_{00}$ | Offset |
| $Y_{10}$ | Hall sensitivity proportional to $B_z$ |
| $Y_{22}$ | Planar hall contribution, proportional to the product $B_x$, $B_y$ |
| $Y_{32}$ | '3D Hall effect', proportional to the product $B_x$, $B_y$, $B_z$ |

To perform the rotation, a device called 'the calibrator', using three pick-up orthogonal coils (coil 1, 2, 3) placed on a rotating platform to measure each component of magnetic field, was built [38]. In the coil support plate, one or several *B*-sensor cards containing the Hall sensor to be calibrated are fixed with a dowel pin (see Fig. 33). The *B*-sensor card, build by NIKHEF Amsterdam, contained in that case three Siemens KSY44 single-axis probes glued on a glass cube and all the electronics, in particular a 24-bit ADC. The *B*-sensor cards were rotated continuously and slowly in a constant and homogeneous field. It was produced either by a magnet at CERN reaching 1.4 T, or for magnetic fields up to 2.5 T by a 130 mm-aperture solenoid at the High Magnetic Field Laboratory of Grenoble (France). The reference value of the magnetic field was obtained from a NMR probe and the angle of rotation measured with a precision of 0.02 mrad. The temperature is stabilized at level of ± 0.02°C. The Hall probes, calibrated with this procedure, successfully mapped the axial magnetic field of the 2 T solenoid of the ATLAS experiment at the LHC. More details about the reconstruction parameters and the results can be found in Ref. [39]. The recent improvements performed on the calibrator are presented in Ref. [40]. However, the major difficulty remains the size of the apparatus. A suitable calibration magnet should have a large aperture or gap (~100 mm), a good homogeneity on a large scale, and produce a high magnetic field. The use of this technique at the level of mass production implies

- the construction of 'light' calibrators equipped with miniaturized sensor cards;
- simplified electronics and reliable calibration software that reconstructs coherent calibration parameters — usually non-linear — based on all these measurements.

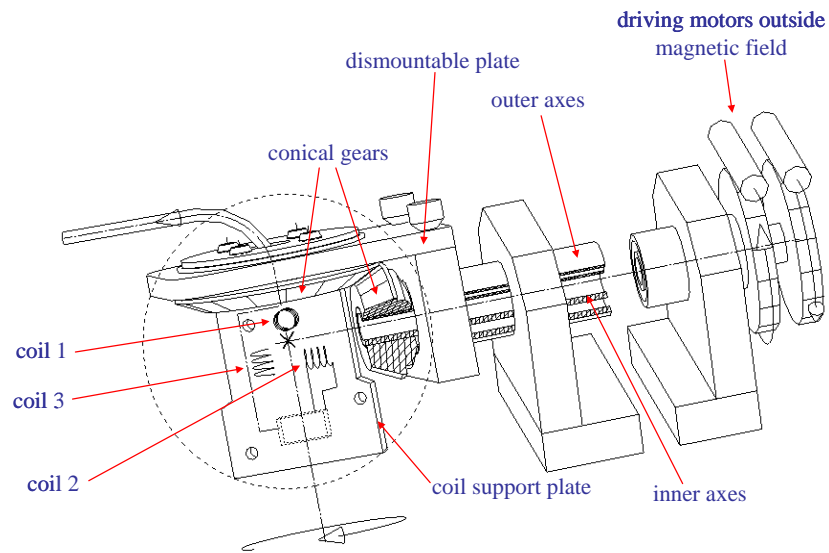

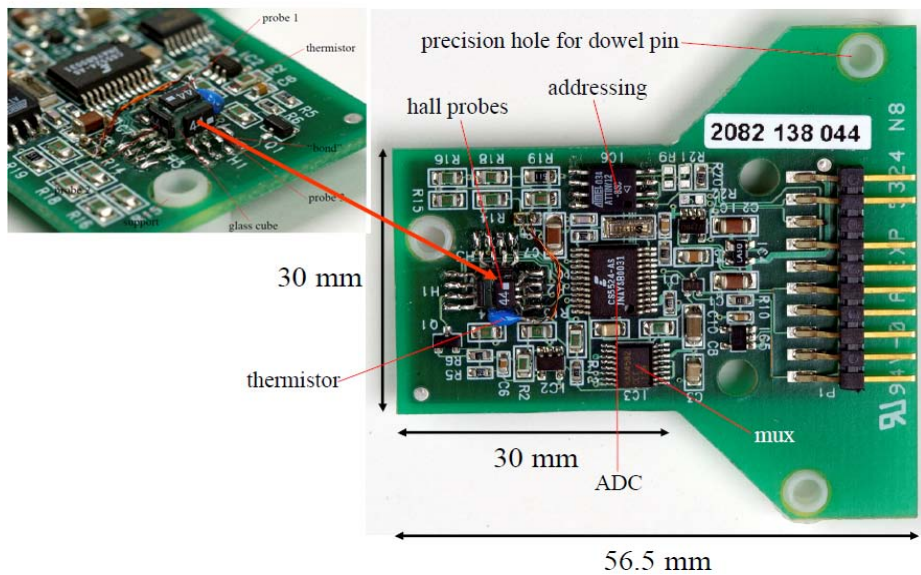

**Fig. 33:** Calibrator (top) and *B*-sensor card built by NIKHEF Amsterdam (bottom) containing the Siemens KSY44 probes glued onto a glass cube [38]

# 6 Application to magnetometry

## 6.1 Field measurement accuracy

The best method for measuring a magnetic field has to be selected after balancing the requirements of the characteristics of the magnetic field to be mapped [41]. The following parameters have to be considered:

- the quantity to measure, e.g., field components, total field, integral, gradient;
- the field measurement range;
- the reproducibility and accuracy;
- the mapped volume and field geometry;
- the time bandwidth;
- the environnent (air, vacuum, cryogenic);
- the use, e.g., research and development, production.

Figure 34 reports a summary of the capabilities of the various methods to measure a magnetic field as discussed in the lectures by L. Bottura and K.N. Henrichsen [16, 42], in terms of measurement accuracy and measurement range. The measurement with Hall probes is particularly suitable for a broad range of magnetic fields up to 10 T or more. The absolute accuracy of the measurement is affected by the parasitic effects discussed in Section 2.4: non-linearity, temperature dependence, alignment errors, and the stability of the offset. Using a reproducible and periodic calibration, a control of the temperature and various compensation techniques, an accuracy value of $10^{-4}$ can be routinely achieved with a single-axis Hall probe. Commercial multi-axis sensors achieve routinely an accuracy of $10^{-2}$. Precautions in the calibration and techniques to reduce the noise and the offset drifts have to be applied to approach the level of 100 ppm.

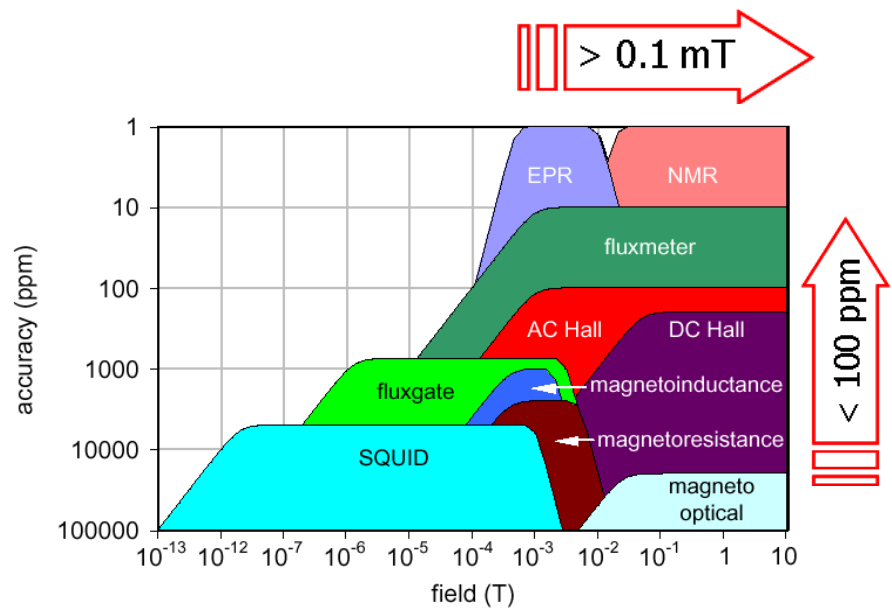

**Fig. 34:** Accuracy of magnetic field measurements performed with different techniques as a function of the magnetic field level [16, 42]

Going more into details, Table 4 summarizes a list of advantages (+) and drawbacks (-) in magnetic field measurements using a Hall effect device.

**Table 4:** Magnetometry using a Hall device, advantages and drawbacks

| (+) | (−) |
|---|---|
| ▪ Easy to use, easily portable/ moved<br>▪ Particularly suited for complex geometries<br>▪ Inexpensive, big market available<br>▪ Fast measurement (instantaneous response<br>▪ Cover a broad range of magnetic fields<br>▪ Medium accuracy for single component measurement (~0.01%), resolution ~0.5 G<br>▪ It is a field mapper: it can measure the three components<br>▪ Can be used for time varying magnetic fields<br>▪ Works in non-uniform fields<br>▪ Works at cryogenic temperatures | ▪ Temperature sensitivity<br>▪ Non-linearity $V_H = f(B)$<br>▪ Offset<br>▪ Drift of the sensitivity with time<br>▪ Noise<br>▪ Lower accuracy for multi-axis integrated circuit Hall sensor: 1% to 0.1%<br>▪ Cross-talk between axes (planar Hall effect, higher order of perturbations)<br>▪ Calibration (delicate for multi-axis sensors)<br>▪ Quantum Hall effect (cryogenic temperature) |

## 6.2 Magnetometry in room-temperature beam line magnets

Hall probes are particularly adapted for making:

- point like measurements,
- magnetic field integrals,
- field maps.

As an illustration, the construction of a field map performed at the Paul Scherrer Institut (PSI) in a resistive beam line dipole is briefly presented. The measurement system consists of an automated five-axis positioning bench adaptable with different types of Hall probes (transverse, axial). The probes are positioned such as to measure the main component of the magnetic field. Highly linear Siemens SBV 601-S1 probes with an active area of 2.6 mm$^2$ and a sensitivity of 0.06 V/T measure the component of the magnetic field perpendicular to the active region with an accuracy of 0.1 G. Table 5 summarizes the performance of the system.

**Table 5:** Field measurements using the Siemens SBV 601-S1 Hall probes

| | |
|---|---|
| Hall Probe | Siemens SVB 601S1 |
| Semicond. material | InAs |
| I max | 400 mA |
| U$_{Hall}$ | 60 mV@1T |
| Longitudinal range | 2100 mm |
| Horizontal range | 650 mm |
| Vertical range | 360 mm |
| Long./Transv./Vert. Resolution | 10 mm |
| Maximum calibrated Field | 3.1 T |
| Hall Probe absolute accuracy | 100 ppm |
| Hall probe resolution | 1 µT |
| Temperature sensibility | 70 ppm/°C |

An overview of the measurement system with a picture of a Hall probe is shown in Fig. 35. The probe is attached to a measuring titanium arm that can move with five degrees of freedom (three directions of translation and rotation around two axes). The device slides on compressed air pads over a flat, precise, machine slide granite block.

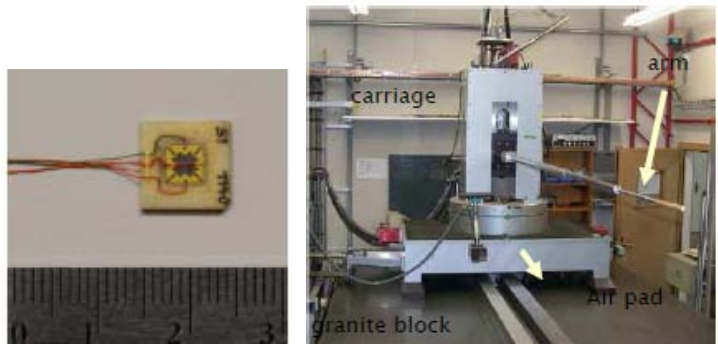

**Fig. 35:** Left: Single-axis Hall probe. Right: Hall probe measurement system developed at the Paul Scherrer Institut

The measurements are performed in a flying mode at a speed of 50 mm/s. The system is placed in a temperature controlled hutch (±0.1°C). The voltages are recorded using HP/Agilent digital multimeters interfaced to a computer via a GPIB gateway. The data are post-processed off line using the calibration procedure described in Section 5.1. For the analysis of a beam line magnet, the knowledge of the main field component is often enough. In the magnetic mid-plane of dipoles, quadrupoles, sextupoles and solenoids, the field mapping of the main component in the magnetic mid-plane provides information like:

– effective magnetic length,
– edge angles,
– edge curvatures,
– beam bending angle,
– beam focusing strength,
– beam vertex position,
– magnetic axis.

Figure 36 shows the results for the main component measured in the mid-plane of a 90 bending magnet. It has been measured from each end and the field maps are here rotated with appropriate angle to each other and put together on the same plot. The magnet outlines are also plotted in thick lines.

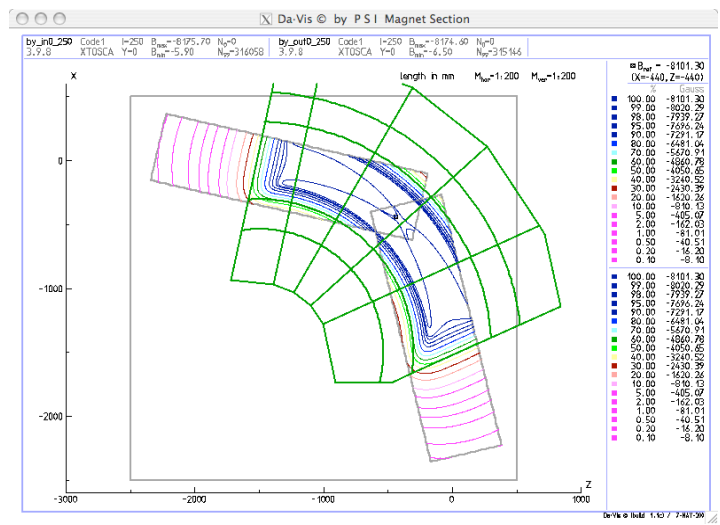

**Fig. 36:** Top view of the field map (main component)

The other field components off the mid-plane are mathematically describable. If necessary, and provided there is good accuracy of the measured main field component, the other two field components could be reasonably well calculated. Two methods are used for creating the full field map in a volume, by extrapolation of the mid-plane fields (already used in other programs like TRANSPORT/TURTLE) and by interpolation between the fields in the mid-plane and two other planes (method developed by the PSI Magnet Section). The extrapolation method is based on the Taylor expansion, therefore relying on high-order derivatives, i.e., the differentiations of the mid-plane fields. All field components off the mid-plane are calculated and the measurement error propagates fast with the distance from the mid-plane. The interpolation method is less sensitive to measurement error and produces more accurate results. The main field component is approximated with an even-order function given by

$$By(x,y,z) = B_0(x,z) + B_2(x,z)\ y^2 + B_4(x,z)\ y^4\ ; (y = 0 \text{ is the mid-plane}). \tag{30}$$

This equation is solved for every ($x,z$) point from the main field component $B_y$ measured in three planes yielding the coefficients $B_0$ (equal to the measured field), $B_2$ and $B_4$. The scalar potential at any point in space is described by

$$V(x,y,z) = B_0(x,z)\ y + B_2(x,z)\ y^3/3 + B_4(x,z)\ y^5/5. \tag{31}$$

The first derivative of potential in any direction is the field component in that direction, thus containing the calculation error to the lowest possible differentiation.

The $B_x$ and $B_z$ field components calculated with the interpolation method on a plane $y = 5$ cm are shown in Fig. 37 (left and right).

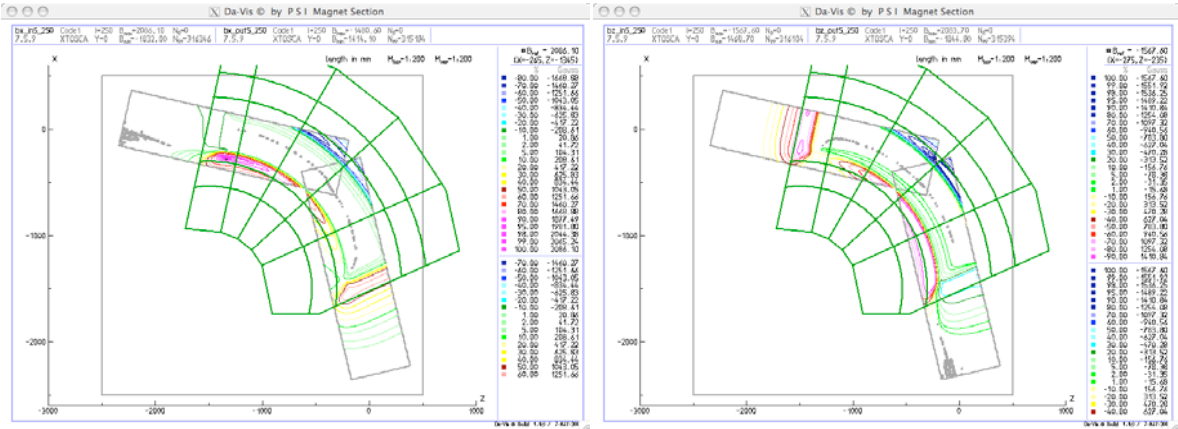

**Fig. 37:** Transverse field components, $B_x$ (left), $B_z$ (right)

### 6.3 Hall probe measurements in undulators

#### *6.3.1 Introduction*

Undulators are powerful generators of synchrotron radiation at storage rings. They are also one of the key components in the production of laser-like radiation down to the angstrom wavelength regime with X-ray free electron lasers. Details concerning the science and the technology of undulators can be found in Refs. [B10, B11, B12] and are also presented in these proceedings [43]. We recall briefly the minimum of information to help in their understanding. Insertion devices can generate a sinusoidal planar field with high peak intensity (up to 2 T) and short period using a periodic magnetic field produced by arrays of magnets of alternating polarity. The electrons 'wiggle' back and forth and the acceleration produced causes the emission of radiation at each pole (see Fig. 38).

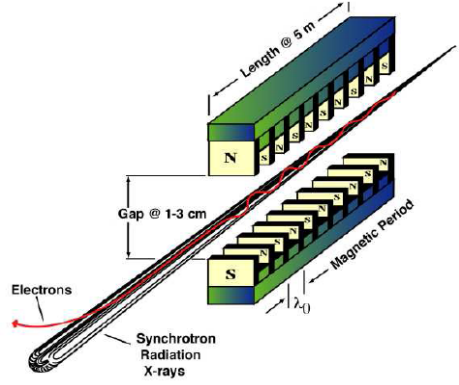

**Fig. 38:** Schematic of a planar undulator

In the case of undulators, the light emitted at each pole interferes with that emitted from others and a quasi-monochromatic spectrum will be observed. At a resonance wavelength λ given by Eq. (32), interference occurs because the emitted light overcomes the electrons by exactly one wavelength. λ is related to the geometric period length $\lambda_o$ of the dipole arrangement but shrunk by $1/2\gamma^2$ ($\gamma = E/E_0$ is the Lorentz factor). This reduction is caused by the conjunction of two relativistic effects. A first shrinking factor of $1/\gamma$ is due to the Lorentz contraction, the electrons see the Lorentz contracted period. Another $1/2\gamma$ results from the Doppler effect when the radiation emitted in the reference frame of the electrons is observed in the laboratory frame. The produced wavelength depends on the period and the energy, but also on the deflection parameter $K$ defined in Eq. (32) and on the observation angle $\Theta$. The number $n$ is the harmonic number of the emitted radiation ($n = 1; 3; 5; \ldots$) at the undulator axis.

$$\lambda = \frac{\lambda_0}{2n\gamma^2}(1 + K^2/2 + \gamma^2\Theta^2)  \qquad (32)$$

$$K = 0.0934\, B_0\, \lambda_0 \qquad (B_0 \text{ is in T, } \lambda_0 \text{ in mm}). \qquad (33)$$

Here $K$ is related to the magnet periodicity and to $B_0$, the peak magnetic field on the undulator axis. As shown by Eqs. (32) and (33), $\lambda$ can be tuned by adjusting the magnetic field. This adjustment is done by changing the magnetic gap between the lower and upper arrays of the magnet blocks (see Fig. 39). When the undulator is linearly polarized, electrons wiggle in one plane. Undulators can also be elliptically polarized, so electrons travel in helixes generating elliptically polarized light. The case of $K_z = K_x$ corresponds to a circular polarized light.

Technologies used to build the magnets for the undulator are:

– The Pure Permanent Magnet (PPM) technology:
  Magnets are made of NdFeB (remanent field $B_r = 1.2$–$1.4$ T) or of $Sm_2Co_{17}$ ($B_r = 1.05$ T). To generate a sinusoidal field, two arrays of permanent magnets should ideally be positioned with the easy axis rotating through 360° per period along the direction of the electron beam. In practice, a good approximation is done by splitting the system into rectangular magnet blocks, M per period (see Fig. 39).
– The Hybrid technology:
  The vertical magnetized magnets are replaced by high permanent iron poles to increase the on-axis flux density.
– The electromagnet technology for long period undulators.
– The superconducting technology with NbTi as superconductor, for fields higher than 2 T.

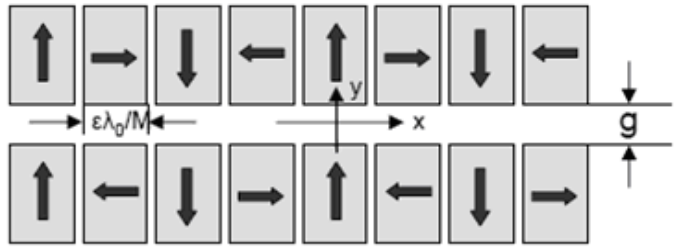

**Fig.39:** Approximation of a permanent magnet undulator by splitting the system into rectangular magnet blocks, four magnet blocks per period (from Ref. [43])

### 6.3.2 *Magnetic field errors in permanent magnet undulators*

The maximum field strength and the characteristics in terms of field errors are figures of merit of the undulator (for a given period length and a given gap). Field errors from permanent magnet undulators are not iron dominated. They come from:

– machining tolerances,
– inhomogeneities of magnetization within the magnet block,
– error in positioning and orientation of the blocks.

Field errors are divided into two categories: the field integral errors and the phase errors.

Knowledge of the first and the second field integral (see formulas in Fig. 40) is essential but not sufficient to characterize the magnetic field in the undulator. Measurements of these two quantities give information of the total change in angle and in position of the beam trajectory at the exit of the device. The ideal undulator is transparent to the electron beam, the first and second field integrals in the undulator vanish with zero errors in the block-to-block magnetization and block positioning. In practice, for a better beam stability, it is important to tune the undulator to smaller values. The integral measurements are usually performed using flipping coils, moving or pulse wire techniques, see Refs. [44, 45].

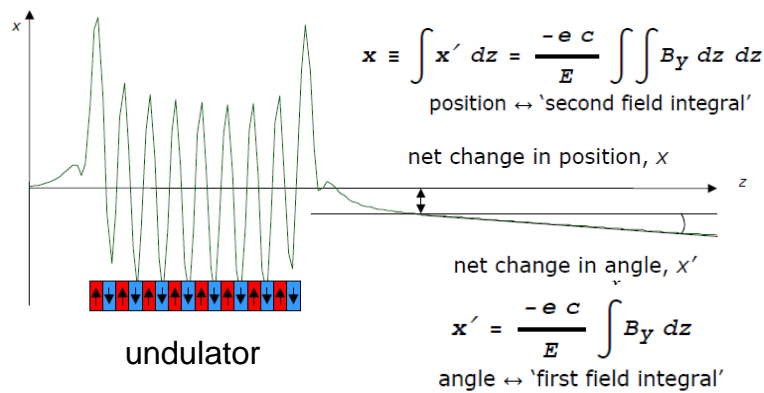

**Fig. 40:** First and second field integral measured along an undulator producing a magnetic field in the $y$ direction (from Ref. [44]). The first field integral gives information on the angle and the second field integral on the position of the electron at the exit of the undulator

Knowledge of the local shape of the magnetic field along a straight line close to the electron trajectory is also important. It provides information about another class of error, the phase error. The phase error indicates the deviation from perfect matching in phase between the electron and the emitted radiation from pole to pole. The slippage between the electron and the light is exactly one period of the emitted radiation ($\lambda$) when one period of the undulator ($\lambda_0$) has been traversed for an ideal magnetic field, i.e., the electron falls behind the light by one period ($\lambda$). A phase error of ±1 degree, for instance, indicates that there is an offset by ±1/360 from one period to the next. It can be characterized as the error in the periodicity of the produced electrical field (see Fig. 41) and has to be kept small to avoid an incomplete constructive interference. A definition of the phase error can be found in Ref. [46]. It is related to the square of the first field integral via the following formula:

$$\varphi(z) = \frac{2\pi}{\lambda_0}\left[\frac{1}{(1+K^2/2)}(z + \frac{e^2}{m^2 c^4} J(z)) - z\right], \quad (34)$$

where $J(z)$ is the integral of the square of the first field integral

$$J(z) = \int_0^z \left[\int_0^z B_y(z')\,dz'\right]^2 dz'. \quad (35)$$

The phase error will attenuate and spread the angular spectral flux on the peak of the spectrum. The origins of this field imperfection are the magnetic peak field fluctuations, the period and the field shape fluctuations. These errors are deduced from local magnetic measurements, performed most widely with a Hall device. After being measured, this error is corrected by shimming.

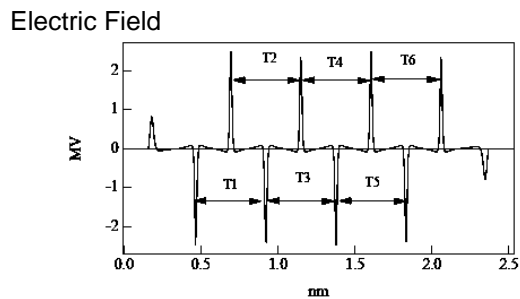

**Fig. 41:** Effect of the phase error observed in the horizontal electrical field produced on-axis by a single electron (from Ref. [47])

### 6.3.3 *Specificity of magnetic measurements in permanent magnet undulators*

The choice of the Hall sensor, the design of the Hall probe bench, and the measurement process have to be adapted in the case of undulator measurements, see Refs. [47, B11]. Insertion devices specifically feature a small gap (several millimetres), a great length (several metres) and a field shape with large longitudinal gradient. This results in the following remarks:

- The measuring device has to be sufficiently small to be inserted in undulators with a small gap. Hall probes with a small size are also preferred for undulators with very short periods. A drastic size reduction will affect the sensitivity of the Hall element.
- On-the-fly scanning is essential to reduce the sensor vibration and the measurement time.
- Temperature stabilization is not critical if the temperature does not change during the measurement over the length of the undulator.
- A high level of accuracy in the Hall probe position on the longitudinal axis is critical, in particular for undulators producing a strong magnetic field gradient.
- The measurement of the three magnetic field components is useful and sometimes required. The measurement accuracy will, however, be limited by the planar Hall contribution. In a device with a vertical field, the determination of the horizontal field integral from numerical integration of the local field can be largely offset. The error is larger in the case of helical undulators. A three-dimensional sensor with a low cross-coupling between the axes (small planar coefficient) should be selected and compensation techniques have to be applied. Investigations on planar Hall effect compensation were carried out on the undulators for the FEL project at the Advance Photon Source (Argonne National Laboratory, USA). Horizontal magnetic field measurements in presence of a vertical field component with a strong gradient were performed [48], using a two-axis Sentron Hall probe based on vertical devices. It was also reported that the planar Hall voltage can be successfully cancelled using a two-sensor device, one Hall probe positioned at 90° with respect to the other [49]. Cancellation techniques like the switching of the voltage/current terminals explained in Section 4.4.4 can be used but will require two undulator scans. At the ESRF (Grenoble, France), the calibration of the Hall probe includes the determination of the planar Hall coefficient, by measuring the three components of the field at various positions in known field geometry of the undulator [50]. The Apple 2 undulators of PETRA 3 (DESY, Germany) were measured using a three-dimensional IC Hall sensor [43].

### 6.3.4  *Hall probe measurements in undulators at the Paul Scherrer Institut*

An example of local field measurements performed in undulators of the Swiss Light Source at the Paul Scherrer Institut is presented for illustration. The design of the Hall probe bench was performed at the ESRF [51]. As displayed in Fig. 42, the bench is based on a granite support mounted on vertically adjustable feet featuring a cross section of 350 x 600 mm$^2$ (width, height). The measuring bench is 4 m long and has an effective length capable of hosting insertion devices with up to 3.5 m magnetic length. The upper granite surface has a flatness of 15 μm and is equipped with guide rails assembled with a parallelism of 5 μm. An Anorad linear motor equipped with a Heidenhain encoder is directly mounted on the granite. The longitudinal position is read out via an optical linear encoder calibrated with a laser interferometer. A carriage is mounted on the guide rails and includes two transverse stages with a travel range of 250 mm. The sensor consists of three one-dimensional Bell Hall elements mounted on a print board (see Fig. 42), fully calibrated up to 1.8 T with a NMR reference magnet. The probes feature a sensitivity of 1 T/V and a residual non-linearity of 0.05% after correction. The bias current is 5 mA. The Hall voltage is read by three Keithley precision voltmeters and the output noise for a 20 ms integration time is less than 0.06 G. They measure all three components of the magnetic field along the aperture of the insertion device under test with an accuracy of $2 \times 10^{-4}$ T. The sensor position is known within 20 μm after software correction based on laser calibration. The functionalities of the measuring bench were implemented in C and C++ with an interface made in Igor PRO. The magnetic field measurements are post-processed with the SRW code developed at ESRF [52].

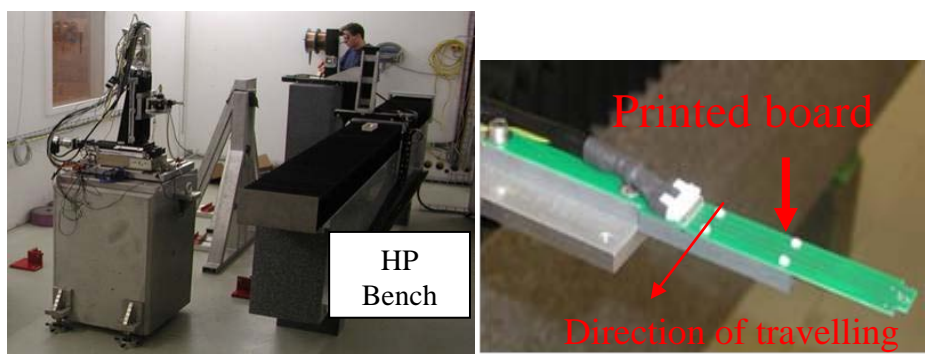

**Fig. 42:** Left: View of the Hall probe bench. Right: Hall probe keeper holding the printed circuit with the Hall elements

The Hall probes travel along the undulator at a typical speed of 10 mm/s. Some 2500–5000 points are recorded. The typical shape of the magnetic field as a function of the longitudinal position measured in a planar undulator is presented in Fig. 43. The first field integral is calculated from the data with an accuracy of 0.5 Gm but has to be calibrated with the result obtained by another measuring method because of the non-linearity of the probes. This measurement aims to investigate the existence of random local angular 'kicks' along the undulator axis. The post-processing of the data results in the measurement of the phase error along the undulator axis (see example in Fig. 43).

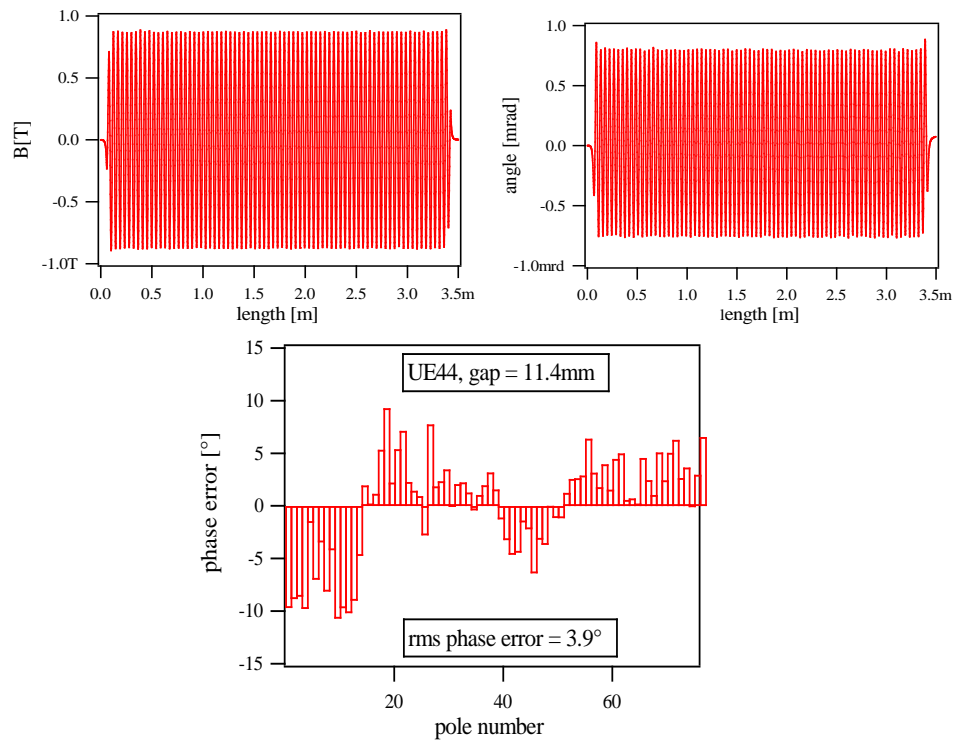

**Fig. 43:** The first integral and the longitudinal variation phase error can be calculated from the measured field along the longitudinal direction

## 7   Conclusions

This lecture reviewed the properties of Hall generators and their use in magnetic field measurements. Hall probes are particularly adapted for point, high speed, and large magnetic field volume measurements. It also offers the advantage of using simple measuring arrangements and low costs. Single-axis Hall probes typically offer an accuracy in the range of 100 ppm. They can be used at cryogenic temperatures and at high magnetic fields. Drawbacks are the need for periodical calibrations and for temperature control, and the existence of parasitic contributions like the planar Hall voltages that have to be compensated. Considerable improvements have been made to reduce the size of the magneto-transducers and develop relatively easy-to-use systems to measure all three magnetic field components in a small spot. The evolution brought by the integration of electronics in vertical (silicon) Hall elements results in the development of the so-called integrated circuit Hall sensors, capable of performing a 3D field mapping in a sub-millimetre size. Techniques have been developed to minimize the offset value and its drift: the planar Hall effect and the 1/$f$ noise. The development of these multi-axis micro-sensors has extended the application of Hall effect sensors, for example, in domains like geomagnetism, spatial positions of objects, contactless automation. A new generation of Hall magnetometers based on IC sensors is now available, see Refs. [29, 30, 41, 53]. They are compact and flexible and are developed with standardized interfaces. For magnetometry, however, the drawback of the three-axis IC sensor is the limited accuracy w.r.t. a conventional single-axis Hall probe. The challenges will therefore concern:

- The improvement, by at least one order of magnitude, of the accuracy of such multi-axis sensors. They feature routinely an accuracy between 0.1% and 1% of the measurement range.
- The ability to perform regular three-dimensional calibrations using standard procedures, interfaces and analysis software.
- The capacity to cover a wider range of magnetic fields, in particular the low field domain for medical applications.
- The lowering of the cost that will enable customers to use three-axis instruments instead of the 'classical' single-axis approach.

## 8    Acknowledgements

The author would like to thank V. Vrankovic and T. Schmidt for their contribution to this lecture. The author is grateful to P. Keller (Metrolab Instruments Geneva), to D. Popovic-Renella and R. S. Popovic (Senis Gmb Zürich) for the valuable discussions concerning the three-axis Hall sensors and for the products they allowed me to show as examples during the lecture. The author is also grateful to D. George and M. Calvi for having the patience to correct this document.

**Bibliography**

Many books refer to the Hall effect device. Those of this list were used to prepare this lecture. A large amount of the information and pictures in this paper originates from Ref. [B1].